\documentclass[12pt]{article}
\usepackage[headsepline]{scrlayer-scrpage}
\usepackage{catchfilebetweentags}
\usepackage{etex}
\usepackage{setspace}
\usepackage[a4paper, left=3.5cm, top=3.5cm, bottom=2cm, right=3cm,foot = 1cm,showframe = false,head = 1cm]{geometry} 
\usepackage[latin1]{inputenc} 
\usepackage{enumitem}
\usepackage{textcomp} 
\usepackage{pdflscape}
\usepackage{graphicx} 
\usepackage{diagbox}
\usepackage{longtable}
\usepackage{flafter} 
\usepackage{amsmath}
\usepackage{multirow}
\usepackage{dsfont}
\numberwithin{equation}{section}
\numberwithin{figure}{section}
\numberwithin{table}{section}
\usepackage{mathtools}
\usepackage{amssymb}
\usepackage{amsfonts}
\usepackage{amsthm}
\usepackage{bm}
\usepackage{placeins}
\usepackage{caption, booktabs} 
\usepackage[natbibapa]{apacite}
\usepackage{url}
\usepackage[pdfencoding=auto, psdextra, hidelinks]{hyperref}
\usepackage{pdflscape}
\usepackage{lmodern}
\usepackage{pdfpages}		
\usepackage{floatflt}
\usepackage{rotating}
\usepackage{pdfsync} 
\usepackage[T1]{fontenc}
\usepackage{tikz}
\usetikzlibrary{decorations.pathreplacing}
\usetikzlibrary{shapes,fit,arrows,positioning}
\usetikzlibrary{shapes.multipart}
\usetikzlibrary{matrix}
\usetikzlibrary{calc}
\tikzstyle{mybox}=[draw=black!80!white,fill=gray!5, thick,
rectangle ,rounded corners]
\tikzstyle{l} = [draw, -latex,thick,black]
\everymath{\displaystyle}

\usepackage{float}
\allowdisplaybreaks[1]
\newcommand\fo[1]{}

\DeclareMathOperator{\rank}{rank}

\DeclareMathOperator{\tr}{tr}
\DeclareMathOperator{\diag}{diag}

\newcommand\norm[1]{\left\lVert#1\right\rVert}

\newcommand\numberthis{\addtocounter{equation}{1}\tag{\theequation}}

\newcommand{\T}{^{\prime}}

\def\mat[#1#2]#3{\left#1\vcenter{\halign{

\hfil$##$ \hfil & \hfil$##$ \hfil & \hfil$##$ \hfil & \hfil$##$ \hfil & \hfil$##$ \hfil & \hfil$##$ \hfil & \hfil$##$ \hfil & \hfil$##$ \hfil & \hfil$##$ \hfil & \hfil$##$ \hfil & \hfil$##$ \hfil & \hfil$##$ \hfil & \hfil$##$ \hfil & \hfil$##$ \hfil & \hfil$##$ \hfil & \hfil$##$ \hfil & \hfil$##$ \hfil & \hfil$##$ \hfil & \hfil$##$ \hfil & \hfil$##$ \hfil & \hfil$##$ \hfil & \hfil$##$ \hfil & \hfil$##$ \hfil & \hfil$##$ \hfil & \hfil$##$ \hfil & \hfil$##$ \hfil & \hfil$##$ \hfil & \hfil$##$ \hfil 

\cr \matA#3;;}}\right#2}
\def\matA#1;{\if;#1;\else #1\cr \expandafter \matA \fi}

\automark[section]{section}
\ofoot{\pagemark}

\begin{document}
	\begin{titlepage}
		
		\title{\large{\textsc{Sequential Estimation of Multivariate Factor Stochastic Volatility Models}}}
		\author{{\normalsize Giorgio \textsc{Calzolari}${}^{1}$ and Roxana \textsc{Halbleib}${}^{2}$ and Christian \textsc{M\"ucher}${}^{3,4}$}}
		\date{\today}
		
		\maketitle
		
		\thispagestyle{empty}

		\begin{abstract}

			\vspace{0.1 cm}
\noindent We provide a simple method to estimate the parameters of multivariate stochastic volatility models with latent factor structures. These models are very useful as they alleviate the standard curse of dimensionality, allowing the number of parameters to increase only linearly with the number of the return series. Although theoretically very appealing, these models have only found limited practical application due to huge computational burdens. Our estimation method is simple in implementation as it consists of two steps: first, we estimate the loadings and the unconditional variances by maximum likelihood, and then we use the efficient method of moments to estimate the parameters of the stochastic volatility structure with GARCH as an auxiliary model. In a comprehensive Monte Carlo study we show the good performance of our method to estimate the parameters of interest accurately. The simulation study and an application to real vectors of daily returns of dimensions up to 148 show the method's computation advantage over the existing estimation procedures. 
			
			\vspace{1 cm}
			
		\end{abstract}

		\thispagestyle{empty}

		\footnotetext[1]{University of Firenze; email: giorgio.calzolari@unifi.it}
		\footnotetext[2]{University of Freiburg; email: roxana.halbleib@vwl.uni-freiburg.de}
		\footnotetext[3]{Corresponding author. Chair of Statistics and Econometrics, Institute of Economics, University of Freiburg, Rempartstr. 16, 79098, Freiburg, Germany; University of Freiburg, email: christian.muecher@vwl.uni-freiburg.de, telephone: +49 761 203-2341}
		\footnotetext[4]{Graduate School of Decision Sciences, University of Konstanz}

	\end{titlepage}

\newpage
\normalsize
\onehalfspacing
\setcounter{page}{1}
\section{Introduction}
Modeling and forecasting the multivariate volatility of financial returns is crucial for risk and portfolio management. Most investors hold large baskets of financial assets whose risks are time-varying and correlated in time. Dynamic approaches to capture, within a multivariate framework, the unobserved time-varying variation and correlation of the financial returns include the multivariate GARCH (MGARCH) models and the multivariate stochastic volatility (MSV) models. While the GARCH specifications are relatively restrictive as they treat the (co-) variance as conditionally deterministic, stochastic volatility models are more flexible as they allow for the variances and correlations to be stochastic. However, the MSV models are difficult to estimate since the (co-) variances and their noises are latent. Moreover, due to the curse of dimensionality, since the number of parameters increases at least quadratically in the dimension of the return vector, MSV models have found only limited application in practice.

A solution to the curse of dimensionality of MSV is to impose a factor structure on the vector of underlying returns, where the factors and idiosyncratic noises follow independent autoregressive stochastic volatility (ARSV) processes. This is known as the multivariate factor stochastic volatility (MFSV) model and has been introduced by \cite{harvey94} (further developed by \cite{pitt99}). This factor representation substantially improves the feasibility of the MSV model in practice by significantly reducing the number of parameters that now increases only linearly with the number of series. However, MFSV still suffers from practical ineffectiveness, even under an exact factor structure, given that the computational burden is enhanced by the latency of the factors, additionally to the variances and idiosyncratic noises.

In this paper, we propose a simple method to estimate the parameters of MFSV models, which allows them to be easily applied in practice, even to very large dimensions and at little computational costs. It consists of two steps: in the first step, we estimate the parameters describing the factor structure, i.e., the loadings and the unconditional variances of the factors and of the idiosyncratic noises, by employing the maximum likelihood (ML) to a static factor representation and by using the convergence results of \cite{anderson03}, \cite{bai12} and \cite{bai16} derived for exact factor models with arbitrary dynamics. In the second step, we apply the Efficient Method of Moments (EMM) of \cite{bansal94} and \cite{gallant96} to estimate the ARSV parameters by implementing simple univariate GARCH auxiliary models to each of the extracted static factors and noises from the first step.

This sequential estimation procedure, which has already been successfully applied by \cite{sentana08}, \cite{calzolari13}, \cite{halbleib18} and \cite{halbleib21} in other contexts, is computationally feasible and simple in terms of implementation and running time regardless of the dimension of the return vectors. To the best of our knowledge, it is also the first frequentist procedure to estimate MFSV models, as all existing ones are exclusively Bayesian. Unlike the Bayesian approaches, this paper focuses solely on estimating MFSV models with fixed parameters and not on filtration and forecasting of the latent variables of the model. We show that our method can be straightforwardly applied to larger dimensions (as well as many factors) compared to the existing literature without imposing any constraints on the parameters and at very low computational costs.

The existing Bayesian procedures, which simultaneously estimate the parameters and filter the latent variables of MFSV, build on the Markov Chain Monte Carlo (MCMC) method as first proposed by \cite{pitt99}. Although theoretically appealing, this method becomes cumbersome when increasing the dimension of the return vector, as it samples from the entire posterior distribution of the model at each iteration. \cite{chib06} partially fix this problem and show that, conditionally on the factor loadings and the factors, the problem reduces to sampling from univariate ARSV processes, which significantly decreases the computational burden of the sampling algorithm and allows for efficient sampling from the posterior distribution also for higher dimensional vectors of returns. Thus, \cite{han05} and \cite{nardari07} apply the sampling algorithm of \cite{chib06} to vectors of around 30 returns and three factors and to vectors of 10 returns and five factors, respectively. \cite{kastner17} improves further the efficiency of the MCMC algorithm proposed by \cite{chib06}, making it computationally faster, but still burdensome in terms of memory requirements, as it involves storing draws from a very high dimensional joint density function of all parameters and of all latent variables (factors, noises, and their respective variances) at each time observation. Thus, the procedure of \cite{kastner17} becomes computationally cumbersome for large vector dimensions, many factors, and many observations. For this reason, the algorithm is applied to only up to 26 series and 5 factors. \cite{kastner19} makes use of Bayesian shrinkage to constrain the dimension of the loading parameters in order to apply the procedure to larger dimensions, such as 300 returns. Nevertheless, as stated in the original paper, the possible parallelization makes the algorithm computationally more intensive in terms of memory requirements.

We show both in our simulation exercise and empirical application that the frequentist sequential method we propose in this paper can estimate the parameters of the MFSV model for very large dimensions of the return vector within minimal computational time and with no memory constraints, especially when parallelizing the equation-by-equation estimation of the second step of the procedure. In the simulation exercise we show that, even without parallelization, we can obtain accurate estimates for vectors of returns of dimension 10 and 2 factors in less than 13 seconds if feeding starting values close to the true parameter values and less than 82 seconds if the starting values are random. Moreover, we find the optimal number of simulations in the EMM estimation that increases the efficiency of the estimates with no additional computational costs as it is adaptive to the length of the series.

In the empirical application we provide further evidence on the speed of our procedure when estimating MFSV for a vector of 148 returns and up to three factors. Here, we show that, as expected, the computational advantage of our method becomes more pronounced once we use multiple CPU cores to run the second step of the estimation in parallel. With parallelization enabled, the computational advantage in the real data application increases by a factor larger than 200. While in simulations our method converges in seconds, in the empirical application it may take only up to 10 minutes. Thus, the method we propose in this paper is the first that manages to practically estimate large dimensional MFSV models without any (shrinkage) constraints on the parameters. While the filtration of the latent variables does not make the object of this paper, based on the estimates proposed here, it takes in the simulation up to approximately 160 minutes using the Bootstrap Particle Filter of \cite{gordon93}, making it at least ten times faster than the most efficient Bayesian method, which runs roughly 24 hours. In the empirical application the Bootstrap Filter is not applicable due to the high number of latent series \citep{bengtsson08}. We, therefore, apply the Auxiliary Multiple Particle Filter of \cite{muecher23} to the data. The run time of the filter with $10240$ particles per series is approximately 20.5 hours, making it roughly 2.9 times faster than the Bayesian method, which runs more than two days for the empirical exercise.

The remainder of this paper is structured as follows. Section 2 formally introduces the MFSV model. In Section 3 we present the estimation method proposed in this paper. Section 4 presents the results from a comprehensive simulation study, and in Section 5, we present the results from applying our procedure to real data. Section 6 concludes.

\section{Multivariate Factor Stochastic Volatility Model}\label{MFSV}
\cite{harvey94} and \cite{pitt99} introduce factor models in the context of MSV modeling to capture the common dynamics and conditional heteroskedasticity in financial returns and to reduce the curse of dimensionality of standard MSV models.

The MFSV is defined for $\bm{y}_{t}$, a vector of $N$ return series observed at time $t$ such that
\begin{equation}
\bm{y}_{t} = B \bm{f}_{t} + \bm{\epsilon}_{t} \;,\;
\mat[()]{\bm{\epsilon}_{t}; \bm{f}_{t}} \sim \mathcal{N}\left[\mat[()]{\boldsymbol 0_{N\times 1} ; \boldsymbol 0_{k\times 1}} , \mat[()]{\Sigma_{t} & 0_{N\times k}; 0_{k\times N} & \Gamma_{t}}\right] \label{measurement_y}
\end{equation}
where $\bm{f}_{t}$ is a vector of $k\times 1$ latent factors, independent of each other, $B$ is of dimension ${N\times k}$ and contains the factor loadings with $k \leq N$ and $\rank(B) = k$ and $\bm{\epsilon}_{t}$ is a $N\times 1$ vector of idiosyncratic components that are orthogonal to the factors. $\Sigma_{t}$ is the diagonal covariance matrix of $\bm{\epsilon}_{t}$ with the elements $\sigma_{t,1}^{2}, \hdots, \sigma_{t,N}^{2}$. $\Gamma_{t}$ denotes the diagonal covariance matrix of the $k$ factors with the elements $\gamma_{t,1}^{2},\hdots,\gamma_{t,k}^{2}$. For identification, we restrict $B$ such that $b_{jj} = 1 \text{ and } b_{ij} = 0 \, \forall j > i, \hspace{3mm} j = 1,\hdots,k$ and $i=j, \ldots,N$ \citep[among others]{pitt99,chib06,nardari07}.\footnote{\cite{sentana01} show that assuming heteroskedasticity for the factors, one may relax one of these restrictions. However, given that our estimation strategy presented in the next section relies on estimating a static factor model, we need both restrictions on the factor loadings matrix, paired with the restrictions that $\Sigma_{t}$ and $\Gamma_{t}$ are diagonal (also implying that their unconditional counterparts, $\mathbb{V}[\bm{\epsilon}_{t}]$ and $\mathbb{V}[\bm{f}_{t}]$, are diagonal) in order to achieve full identification in the static factor model \citep{bai12,cox17,williams20}.} Thus, the loading parameters left for estimation are $b_{ij}$ with $j=1,\hdots, k$ and $i=j+1,\hdots, N$.

Let $\bm{x}_{t} = (\epsilon_{t,1},\hdots,\epsilon_{t,N},f_{t,1},\hdots,f_{t,k})'$ such that each of the elements of $\bm{x}_{t}$ follows a univariate Autoregressive Stochastic Volatility (ARSV) process as it follows:
\begin{align}
x_{t,m} &= \exp{(h_{t,m}/2)} u_{t,m} & u_{t,m}\overset{i.i.d}{\sim}	\mathcal{N}(0,\;1) \label{measurement_x}\\
h_{t,m} & = \mu_{m} + \varphi_{m}(h_{t-1,m}-\mu_{m}) + \sigma_{\eta,m}\eta_{t,m} & \eta_{t,m} \overset{i.i.d}{\sim} \mathcal{N}(0,1)\label{transition_x}
\end{align}
where 
\begin{equation}h_{t,m} = \begin{cases} \ln{(\sigma_{t,m}^{2})} & \text{for } m = 1,\hdots, N \\ 
\ln{(\gamma_{t,m-N}^{2})} & \text{for } m = N+1,\hdots, N+k, \end{cases}
\end{equation} 
$|\varphi_{m}|<1$ that assures covariance-stationarity and $\sigma_{t,m}^{2}>0$.
Thus, the conditional covariance matrix of the return series at time $t$ is given by:
\begin{equation}
\mathbb{V}[\bm{y}_{t}\vert \bm{h}_{t}] = B\Gamma_{t}B\T + \Sigma_{t},
\end{equation} 
where $\bm{h}_{t} = (h_{t,1},\hdots,h_{t,N+k})\T$ is the vector of log variances of the factors and idiosyncratic errors. The model has a total of $Nk - k(k+1)/2 + 3(N+k)$ parameters that increases linearly in the number of return series $N$ rather than quadratically, as is the case of the MGARCH and general MSV models. However, the estimation of the parameters of the MFSV is difficult since the factors $(f_{t,1},\hdots,f_{t,k})'$, the errors $(\epsilon_{t,1},\hdots,\epsilon_{t,N})'$ and their stochastic variances $(\gamma_{t,1}^{2},\hdots,\gamma_{t,k}^{2},\sigma_{t,1}^{2},\hdots\sigma_{t,N}^{2})'$ are latent. 

\section{Estimation}\label{estimation}
In what follows, we present our simple frequentist approach to estimate the parameters of the MFSV model based on two steps. For this reason, we rewrite the vector of the model parameters as $\bm{\theta} = \left(\bm{\theta}_{1}, \bm{\theta}_{2}\right)\T$, where:
\begin{equation}\bm{\theta}_{1} = \left(b_{21}, \hdots, b_{N1}, b_{32},\hdots b_{N2},\hdots b_{Nk}, \sigma_{1}^{2},\hdots,\sigma_{N}^{2}, \gamma_{1}^{2},\hdots,\gamma_{k}^{2}\right)'
\end{equation}
is estimated in the first step (to be described in Section \ref{part1}) and
\begin{equation}\bm{\theta}_{2} = \left(\varphi_{1},\hdots,\varphi_{N+k},\sigma_{\eta,1}, \hdots,\sigma_{\eta,N+k}\right)'
\end{equation}
is estimated in the second step (to be described in Section \ref{part2}). 

Note that $\bm{\theta}_{2}$ does not contain the ARSV constants $(\mu_{1},\hdots,\mu_{N+k})'$, since we can identify them from the unconditional variances of the ARSV processes as it follows:
\begin{equation}
\mu_{m} = \ln{(\psi_{m})} - \frac{\sigma_{\eta,m}^{2}}{2(1-\varphi_{m}^{2})}, \quad\forall\quad m=1,\ldots, N+k,\label{uncondvarARSV}
\end{equation}
where $(\psi_1,\ldots,\psi_{N+k})'= \left(\sigma_{1}^{2}, \hdots,\sigma_{N}^{2}, \gamma_{1}^{2},\hdots,\gamma_{k}^{2}\right)'\equiv\bm\psi$.

\subsection{Estimation of the Factor Model Parameters by ML}\label{part1}
This section provides the estimates of $\bm{\theta}_{1}$. For this we define a static factor model as it follows:
\begin{align}
\label{eq1}\bm{y}_{t} &= B^{*}\bm{g}_{t} + \bm{e}_{t}, \\
\label{eq2}\bm{e}_{t} &\overset{i.i.d}{\sim} \mathcal{N}\left({\bm 0_{N\times 1}} , \Sigma^{*}\right),\\
\label{eq3} \bm{g}_{t} & \overset{i.i.d}{\sim}\mathcal{N}\left(\bm{0}_{k\times 1},\Gamma^{*}\right),
\end{align}
with the same assumptions and restrictions on $B^{*},\bm{g}_{t}, \Sigma^{*},\Gamma^{*}$ as in the MFSV model presented in Section 2. We achieve full identification of the static factor model from the restrictions on the factor loadings matrix and the unconditional covariance matrices of the static factors $\bm{g}_{t}$ and of the static factor model idiosyncratic errors $\bm{e}_{t}$. For details on the identification see \cite{williams20}. Denote the parameter vector of this static factor model by 
\begin{equation}\bm{\beta}_{1} = (b_{21}^{*},\hdots,b_{Nk}^{*},{\sigma_{1}^{*}}^{2},\hdots,{\sigma_{N}^{*}}^{2},{\gamma_{1}^{*}}^{2},\hdots,{\gamma_{k}^{*}}^{2})\end{equation}
i.e., $\bm{\beta}_{1}$ gives the factor loadings, the unconditional variances of the factors, and the unconditional variances of the idiosyncratic errors of the static factor model defined in equations (\ref{eq1}) - (\ref{eq3}).
\cite{bai12} and \cite{bai16} show that, in the exact factor model case, for $T \to \infty$, \begin{equation}\label{convergence}
\hat{\bm{\beta}_{1}} \overset{p}{\longrightarrow} \bm{\theta}_{1}
\end{equation}
i.e., the static factor model ML estimators converge to their true parameter counterparts in the dynamic factor model.\footnote{\cite{bai16} show that for approximate factor models, the convergence occurs for $N,T \to \infty$.} This result holds if the factors are independent of the error terms, if the factor variances are bounded away from zero (\citealp{bai12}; \citealp{bai16}) and if the distribution of $(\bm{\epsilon}_{t}\T,\bm{f}_{t}\T)\T$ has finite second-order moments \citep{anderson03}, which are also the assumptions of our factor structure. This result simplifies our estimation procedure drastically, especially when applied to vectors of returns and factors of large dimensions. 

However, as shown by \citet{bien11} and \cite{bai12}, the log-likelihood of the static factor model is multi-modal. Thus, classical numerical optimization methods are infeasible for higher dimensional return vectors $\bm{y}_{t}$ as they are likely to get stuck in local minima. In this paper we use the Expectation Maximization (EM) algorithm proposed by \cite{bai12} to estimate the parameters of the static factor model combined with a modified version of the Gradient Descent algorithm of \citet{bien11}, which considerably reduces the computational time (see also \cite{bai16b} and \cite{daniele19} for a successful implementation of this combination to estimate other types of factor models). The detailed description of the estimation procedure at this step is presented in Appendix \ref{appendix1}.

After obtaining $\widehat{B^{*}}$, $\widehat{\Sigma^{*}}$ and $\widehat{\Gamma^{*}}$, we make use of the asymptotic results of \cite{bai12} and \cite{bai16}, and take them as consistent estimates of ${B}$, $\Sigma$ and ${\Gamma}$, i.e. of $\bm{\theta}_{1}$, by simply setting: $\hat{\boldsymbol{\theta}}_{1} = \hat{\boldsymbol{\beta}}_{1}$.

\subsection{Estimation of ARSV Parameters by EMM}\label{part2}
Based on $\widehat{B^{*}}$, $\widehat{\Sigma^{*}}$ and $\widehat{\Gamma^{*}}$ obtained in the first step described above, we extract the static factors by the projection formula of \cite{bai12}:
\begin{equation}
\hat{\bm{g}}_{t} = \underbrace{\left(\widehat{\Gamma^{*}}^{-1} + {\widehat{B^{*}}}\T{\widehat{\Sigma^{*}}}^{-1}\widehat{B^{*}}\right)^{-1}{\widehat{B^{*}}}\T{\widehat{\Sigma^{*}}}^{-1}}_{\Pi^*}(\bm{y}_{t}-\bar{\bm{y}}),\label{projection}
\end{equation}
where $\bar{\bm{y}}$ is the mean of the return series over the whole sample. Define:
\begin{equation}
\hat{\bm{e}}_{t} = \bm{y}_{t} - \widehat{B^{*}}\hat{\bm{g}}_{t}\label{residuals}.
\end{equation} 
and let $\hat{\bm{x}}_{t} = (\hat{e}_{t,1},\hdots,\hat{e}_{t,N}, \hat{g}_{t,1},\hdots,\hat{g}_{t,k})'$. 

At this second step, we estimate the elements of $\bm{\theta}_{2}$ by applying the EMM approach of \cite{bansal94} and \cite{gallant96}. Given the independence among and between the factors and the noises of the factor structure defined in Section \ref{MFSV}, the EMM approach can be simplified by estimating the parameters of the ARSV models equation-by-equation by applying the appropriate univariate auxiliary model to 
each of the series composing $\hat{\bm{x}}_{t}$. We choose here to apply GARCH(1,1) as the auxiliary model, as it has already been successfully implemented for the EMM estimation of univariate ARSV models by \cite{calzolari04} and \cite{monfardini98}, among others.\footnote{As an alternative to GARCH, we also implement the ARMA(1,1) model fitted to the log squared transformation of the elements of $\hat{\bm{x}}_{t}$ \citep{monfardini98}. However, due to the "flatness" in the ARMA(1,1) likelihood for at least one of the elements of $\hat{\bm{x}}_{t}$, we were not able to get reliable estimates and standard errors of the MFSV parameters based on this choice of the auxiliary model. The results are, however, available from the authors upon request.}

In Appendix \ref{appendix2} we provide the detailed description of the steps taken to undergo the EMM estimation of $\bm{\theta}_{2}$. For this we define here below the structure of the GARCH(1,1) that we apply to each of the series composing $\hat{\bm{x}}_{t}$ as it follows:
\begin{eqnarray}
\hat{x}_{t,m}&=&\delta_{t,m}\xi_{t,m},\\
\delta_{t,m}^{2} &=& (1-\alpha_{1,m}-\alpha_{2,m})\hat\psi_{m} + \alpha_{1,m} \hat{x}_{t-1,m}^{2} + \alpha_{2,m}\delta_{t-1,m}^{2}\label{garch},
\end{eqnarray}
where $m=1,\ldots,N, N+1, \ldots, N+k$, $\xi_{t,1},\ldots,\xi_{t,m}$ are independent white noises, $\alpha_{1,m}>0$, $\alpha_{2,m}>0$ and $\alpha_{1,m}+\alpha_{2,m}<1$ $\forall m$ and $\hat\psi_{m}$ is the estimate of $m$-th element of $\bm\psi$ computed from the estimated elements of $\bm{\theta}_{1}$ obtained in the first step. Here, we follow \cite{sentana01} and \cite{sentana08} and identify GARCH constant through a consistent estimator of the unconditional variance of the respective process. We define $\bm{\beta}_{2}=(\alpha_{1,1},\ldots,\alpha_{1,N+k},\alpha_{2,1},\ldots,,\alpha_{2,N+k})'$ containing all parameters for the $N+k$ univariate GARCH(1,1) models described above that are estimated equation-by-equation by means of Pseudo Maximum Likelihood (PML). 
To speed up the estimation at this step, we can parallelize the estimation of $\bm{\theta}_{2} = \left(\varphi_{1},\hdots,\varphi_{N+k},\sigma_{\eta,1}, \hdots,\sigma_{\eta,N+k}\right)'$ in $N+k$ EMM estimation algorithms for each of the parameter pairs $\bm{\theta}_{2,m} = (\varphi_{m},\sigma_{\eta,m})'$ with $m=1,\ldots, N+k$ to which corresponds the auxiliary parameter vector $\bm{\beta}_{2,m}=(\alpha_{1,m},\alpha_{2,m})'$. The EMM estimation provides consistent estimators of $\bm{\theta}_{2}$ under very general conditions that are fulfilled by our model specification. 

\subsection{Variance Covariance Estimation}
Both parts of the estimation procedure we present above involve the estimation of an auxiliary model: i.e., in the first part, the static factor model, in the second part, univariate GARCH models on the extracted static factors from the first part. While the estimates of the auxiliary model of the first part $\hat{\bm{\beta}}_{1}$ are taken to be the final corresponding estimates (of loadings and unconditional variances) of the true dynamic model, the auxiliary estimates of the second part $\hat{\bm{\beta}}_{2}$ help at computing the EMM estimates of $\bm{\theta}_{2}$. However, in order to compute the standard errors of the estimators of the MFSV parameters we propose here, we implement the asymptotic variance-covariance matrix of the EMM estimator given by \citep{gourieroux93}:
\begin{equation}
W(H) = \left(1+\frac{1}{H}\right)\left[\frac{\partial \mathcal{Q}\left(\tilde{\bm{y}}_{t}(\hat{\bm{\theta}}),\hat{\bm{\beta}}\right)\T}{\partial\bm{\theta}}\widehat{\mathcal{I}}(\hat{\bm{\beta}})^{-1}\frac{\partial \mathcal{Q}\left(\tilde{\bm{y}}_{t}(\hat{\bm{\theta}}),\hat{\bm{\beta}}\right)}{\partial\bm{\theta}\T}\right]^{-1},\label{asyvar}
\end{equation}
where $\widehat{\mathcal{I}}(\hat{\bm{\beta}})$ denotes the Fisher information matrix of the auxiliary model and $ \mathcal{Q}\left(\tilde{\bm{y}}_{t}(\hat{\bm{\theta}}),\hat{\bm{\beta}}\right)$ is obtained by stacking vertically together the auxiliary score vectors 
$\mathcal{Q}(\bm{y}_{t};\hat{\bm{\beta}}_{1})$ and $\mathcal{Q}(\hat{\bm{x}_t};\hat{\bm{\beta}}_{2})$, i.e., $\mathcal{Q}(\bm{y}_{t};\hat{\bm{\beta}}_{1},\hat{\bm{\beta}}_{2}) =(\mathcal{Q}(\bm{y}_{t};\hat{\bm{\beta}}_{1})\T,\mathcal{Q}(\hat{\bm{x}_t};\hat{\bm{\beta}}_{2})\T)\T$, where $\mathcal{Q}(\bm{y}_{t};\hat{\bm{\beta}}_{1})$ are obtained by computing the first derivative of the log-likelihood of the static factor model with respect to ${\bm{\beta}}_{1}$ at $\hat{\bm{\beta}}_{1}$. The closed form expression of this score vector is presented in Appendix \ref{appendix3}. $\mathcal{Q}(\hat{\bm{x}_t};\hat{\bm{\beta}}_{2})$ stacks together the $N+k$ GARCH(1,1) scores vectors: $\mathcal{Q}(\hat{\bm{x}_t};\hat{\bm{\beta}}_{2})=(\mathcal{Q}_{1}(\hat{\bm{x}}_{1};\hat{\bm{\beta}}_{2,m}), \ldots,\mathcal{Q}_{N+K}(\hat{\bm{x}}_{N+K};\hat{\bm{\beta}}_{2,N+K}) )'$ defined in Appendix \ref{appendix2}. \footnote{Note that $\mathcal{Q}(\bm{y}_{t};\hat{\bm{\beta}}_{1},\hat{\bm{\beta}}_{2})$ does not contain $\hat{\bm{x}_t}$, since it is a function of $\bm{y}_{t}$ and $\hat{\bm{\beta}}_{1}$} We define $\hat{\bm{\beta}} = (\hat{\bm{\beta}}_{1}\T,\hat{\bm{\beta}}_{2}\T)\T$.

In Equation (\ref{asyvar}), the Fisher information matrix of the auxiliary model, $\widehat{\mathcal{I}}(\hat{\bm{\beta}})$ is not available in complete form, as the parameters $\hat{\bm{\beta}}$ are estimated in two steps. Here we follow \cite{halbleib21} and replace it with a consistent simulation-based estimator: the sample variance-covariance matrix of 1000 independently simulated score vectors of the overall auxiliary model. These score vectors are computed after the last iteration upon convergence. As all components of the auxiliary model score vector are available in closed form, the simulation-based estimation of the Fisher information matrix and the whole variance-covariance matrix $W(H)$ in Equation (\ref{asyvar}) are computationally very fast. 

As one may see from Equation (\ref{asyvar}), $W(H)$ decreases with $H$, the number of the simulated series in the auxiliary EMM estimation. As there is no rule on how to choose $H$, in our empirical application we follow the main stream of the literature, and choose $H$ in such a way that the precision of the estimates improves, but not at very high computational costs (usually $H$ is set to 10). However, in our simulation exercise, we find also an optimal choice of $H$ adapted to the number of observations available for the empirical study.

In order to get the standard errors of the constant terms of our ARSV processes $\hat{\mu}_{1},\hdots,\hat{\mu}_{N+k}$ are computed from the estimates of first step estimation, we use the Delta Method as described below: 
Remember that 
\begin{equation}
\mu_{m} = \ln{(\psi^{*}_{m})} - \frac{\sigma_{\eta,m}^{2}}{2(1-\varphi_{m}^{2})}.
\end{equation}
and denote $\bm{\xi}_{m} = (\psi^{*}_{m},\varphi_{m},\sigma_{\eta,m})\T$. The derivative of $\mu_{m}$ with respect to the parameter vector $\bm{\xi}_{m}$ evaluated at $\hat{\psi}_{m},\hat{\varphi}_{m},\hat{\sigma}^{2}_{\eta,m}$ is given by:
\begin{equation}
\frac{\partial \mu_{m} }{\partial \bm{\xi}_{m}} = \left(\frac{1}{\hat{\psi}^{*}_{m}}\,,\,-\frac{\hat{\varphi}_{m}\hat{\sigma}_{\eta,m}^{2}}{(1-\hat{\varphi}_{m}^{2})^{2}}\,,\,-\frac{\hat{\sigma}_{\eta,m}}{(1-\hat{\varphi}_{m}^{2})}\right)\T.
\end{equation} 
The variance of $\hat{\mu}_{m}$ is then given by 
\begin{equation}
\mathbb{V}[\hat{\mu}_{m}] = \frac{\partial \mu_{m}}{\partial \bm{\xi}_{m}\T}\mathbb{V}[\hat{\bm{\xi}}_{m}]\frac{\partial \mu_{m} }{\partial \bm{\xi}_{m}},
\end{equation}
where $\mathbb{V}[\hat{\bm{\xi}}_{m}]$ is extracted correspondingly from $W(H)$ defined above.


\section{Monte Carlo Simulation}\label{MC}
This section provides results on the statistical properties of the estimates obtained based on the procedure described in the previous section for various choices of $T$, $N$ and $k$ that are empirically relevant. We use $ N = \{10,20,30,100\}$ simulated return series, $k = \{1,2,3\}$ simulated factors to generate $ T = \{1000,4000,10000\}$ observations. We choose the parameter values similarly to \cite{kastner17}:
\begin{minipage}{0.4\textwidth}
	$B = \left(
	\begin{array}
	{ccc}
	1 & 0 & 0 \\
	0.9& 1 & 0 \\
	\vdots & 0.2 & 1 \\
	& \vdots &0.4+\varepsilon\\
	& 	 		& \vdots \\
	&			&	0.7 \\
	& 			& 0.1 \\
	\vdots&	\vdots &	\vdots \\	
	0.1 & 0.8 & 0.4\\
	\end{array}
	\right),$
\end{minipage}
\begin{minipage}{0.4\textwidth}
	\begin{align*}
	(\varphi_1,\hdots,\varphi_N)' &= (0.9,\hdots,0.99)', \\
	(\varphi_{N+1}, \varphi_{N+2}, \varphi_{N+3})' &= (0.99,0.95,0.91)',\\
	(\mu_1,\hdots,\mu_{N})' & = (-2,\hdots,-1.1)',\\
	(\mu_{N+1},\mu_{N+2},\mu_{N+3})'&= (0,0,0)',\\
	(\sigma_{\eta,1},\hdots,\sigma_{\eta,N})' &= (0.6,\hdots,0.15)',\\
	({\sigma}_{\eta,N+1}, {\sigma}_{\eta,N+2},{\sigma}_{\eta,N+3})'&= (0.2,0.3,0.4)'.
	\end{align*}
\end{minipage}\\[1em]
The dots indicate an evenly spaced grid between the first and the last value. Therefore, we choose the elements of the first column of $B$ such that they are evenly spaced between the values $0.9$ and $0.1$; the elements of the second column are chosen such that they are evenly spaced between $0.2$ and $0.8$; the elements of the third column are chosen to be evenly spaced between $0.1$ and $0.7$, such that $\varepsilon=(0.7-0.1)/(N-4)$. For the simulation cases with $k<3$, we choose the first $k$ columns of $B$ and the first $N+k$-th values from each of the parameter vectors ${\bm\varphi}$, ${\bm\mu}$ and ${\bm\sigma}_{\eta}$. For all simulations, we choose $R = 1000$ replications.

For the step size of the Gradient Descent Method of \cite{bien11}, we use the adaptive moments algorithm with $d = 0.005$.\footnote{The estimation is conducted on the bwHPC Cluster with multiple jobs, each using 40 CPU cores (two Octa-core Intel Xeon E5-2670 (Sandy Bridge) Processors with 2.6GHz) in parallel.}

\subsection{Choose starting values and $H$}
We compare two strategies for selecting the starting values for the parameters of the EMM procedure of the second step. The first strategy is to provide user-specified starting values where we choose them to be 20\% smaller than the true ARSV autoregressive parameters and 20\% larger than the ARSV standard deviation parameters. The second strategy is to obtain starting values from the QML approach of \cite{ruiz94} applied to the extracted static factors and residuals, i.e., we apply the Kalman filter to their log squared transformations. This second strategy is especially appealing for empirical applications where the choice of good starting values is not straightforward.

Next, we analyze the effect that $H$, the number of simulation paths in the EMM, has on the estimation results. We first set $H = 10$ as in \cite{calzolari14}, \cite{halbleib18} and \cite{halbleib21}. However, we find that, for small $T$, increasing $H$ to $100$ improves the performance of our procedure, whereas, for large $T$, the improvement in the performance is negligible while the computational burden hugely increases.\footnote{We do not report the results for $H = 100$. They are available from the authors upon request.} To balance this behavior, we follow the idea of \cite{monfardini98} and choose, besides fixing it to the value of 10, $H$ as a function of the sample size, $T$. Specifically, we set $H$ such that $H\cdot T = 10^5$, i.e. $H = 10^{5}/T$. This is mainly motivated by the reduction in the number of outliers of the EMM estimation procedure, as described below. 

\subsection{Loss function}

We summarize the results of the simulation study by means of the Mean Squared Error (MSE) of the estimated parameters $\hat{\bm{\theta}}^{r}$ over the $r = 1,\hdots,R$ MC replications. We compute the MSE of $\hat{\bm{\theta}}$ by
\begin{equation}\label{MSE}
MSE(\hat{\bm{\theta}}) = \frac{1}{(\dim{\bm{\theta}}) R}\sum\limits_{r = 1}^{R}\sum\limits_{i=1}^{(\dim{\bm{\theta}})}\left(\hat{\theta}^{r}_{i} - \theta_{i}\right)^{2},
\end{equation}
i.e., the average squared deviation of the estimated values from the true values. We also report results on the ratio between the empirical and asymptotic standard deviations of the estimated parameters.

\subsection{Outliers}
Some of the replications yield outliers in the second step of the estimation. Such outliers are often characterized by very large estimated ARSV standard deviation parameters or by negative ARSV autoregressive parameters. Not all cases of outliers are accompanied by large values of the EMM distance, and therefore their identification based on it is very sensitive to the distance threshold we choose. Therefore, we choose a less sensitive way to identify outliers based on the values of the estimated parameters in each MC replication. In particular, we discard the replication with at least one of the following outliers. First, the estimated autoregressive parameter is at least ten times smaller than the true value, or it is negative. Second, the absolute value of the constant or standard deviation parameter is at least ten times larger than its true value. \footnote{In fact, we discard replications for which the absolute value of the estimated constant of the factors is greater than 9, since the true value is zero.}Although these types of outliers only affect the estimated values of the parameters of the corresponding series in the second step procedure (and not the estimates of the first step), they also affect all standard errors of the entire vector of parameter estimates due to the way we compute the variance-covariance matrix of the estimates as described in Section \ref{estimation}.\footnote{Alternatively, we could choose not to discard the outliers and present summary results by means of quantile measures, such as median and the interquartile ranges. These results are available from the authors upon request.} 

Tables \ref{GARCH_NC_std_H_10} to \ref{GARCH_NC_KF_H_adapt} in Appendix \ref{appendix4} report the percentage of outliers for different numbers of factors $k$, number of return series $N$, sample size $T$, number of simulations in the EMM step $H$ and starting values. From analyzing the entries of the tables, we see that the choice of $T$ is crucial in generating outliers: while large values of $T$ are accompanied by only a few outliers (for $T=10000$, only up to 0.5\% of all simulated series are discarded), working with series of smaller sample sizes is much more affected by the outliers: e.g., for $T=1000$, large $N$ ($N=100$), small $H$ (e.g., $H=10$) and poor starting values, we need to discard up to 37\% of the replications. However, increasing $H$ (to 100) reduces the number of outliers drastically, e.g., only 10\% are discarded. A further significant reduction is obtained by choosing the QML starting values: for $H=10$ the percentage of discarded replications decreases to 5\%, while for $H=100$, it further reduces to 3.5\%. Also, for $T=4000$, we find that increasing $H$ and choosing the QML starting values significantly reduces the outliers, especially for large $N$. Moreover, by setting $H=10^5/T$ and using QML starting values, we need to discard only up 3.6\% of the replications, regardless of the choice of $T$ and $N$, which is much less than by keeping $H$ fixed (e.g., to $H=10$). Therefore, in our empirical application presented in Section \ref{real} we choose $H=10^5/T$ and use the QML starting values.

\subsection{Simulation results}


Appendix \ref{appendix5} and \ref{appendix6} report simulation results for the whole vector of the MFSV parameters, i.e., $\hat{\bm{\theta}}$ as well as for sub-vectors of it, i.e., for the factor loadings, the unconditional variances and the ARSV parameters, respectively. In Appendix \ref{appendix5} we focus on presenting the results on the ratio between the MC standard deviation of the estimated parameters and their asymptotic standard errors, computed as described in the Section \ref{estimation}, averaged over the components of the corresponding (sub)-vectors. Tables \ref{GARCH_rat_std_H_10} and \ref{GARCH_rat_KF_H_10} present results for fixed $H=10$ and tables \ref{GARCH_rat_std_H_adapt} and \ref{GARCH_rat_KF_H_adapt} report results for $H = 10^{5}/T$. As expected, increasing $H$ in the EMM estimation procedure reduces the asymptotic variance of the estimated parameters. For $H = 10$ and large $T$, the average ratio between the empirical standard deviation of the estimated parameter values over the Monte Carlo replications and the estimated asymptotic standard errors is close to 1, regardless of the choice of the starting values. For lower values of $T$, however, the ratio is smaller than 1 in most cases, indicating that the asymptotic standard errors are too large compared to the Monte Carlo standard deviation. In these cases, the QML-based starting values improve the ratio towards 1. However, by setting $H = 10^{5}/T$, also for smaller values of $T$, the ratios become close to 1.

 Appendix \ref{appendix6} provides detailed numerical results for the MSE of the estimated parameters for the two choices of $H$ and the starting values. From the tables one may see that, as expected, the MSEs of the parameters estimated in the first part of our procedure are unaffected by the choice of the EMM starting values of $H$. However, for the quality of ARSV parameters estimated in the second part of our procedure, both the choice of $H$ and of the starting values is important and exhibits complementary effects for larger $T$s. Thus, choosing QML starting values of $T$ larger than $1000$ provides ARSV estimates with small MSE also in the case of fixed $H=10$.

Generally, the MSE of the parameter vector $\hat{\bm{\theta}}$ decreases when $T$ increases and increases when $k$ increases. These effects are particularly pronounced for smaller values of $N$ and reduce when $N$ increases. Similar results are obtained by \cite{halbleib21} that show that the burden of adding a new factor to the estimation reduces when the underlying number of returns increases, as it provides more information on their commonalities than when increasing $T$.

By looking in detail at each set of parameters, one may see that increasing $N$ has a stronger positive effect on the quality of the ARSV estimates than on the quality of the loading and the unconditional variance estimates. While the quality of the loading estimates is mainly improved by increasing $T$, the quality of the unconditional factor variances also improves by increasing the number of factors. As the constant parameters are identified from the unconditional variance of the error term and the AR and the standard deviation parameters, it inherits their MSE pattern, especially in what regards the dependence to $k$ and $N$. 

\begin{figure}[hbt]
\includegraphics[width = 1\linewidth]{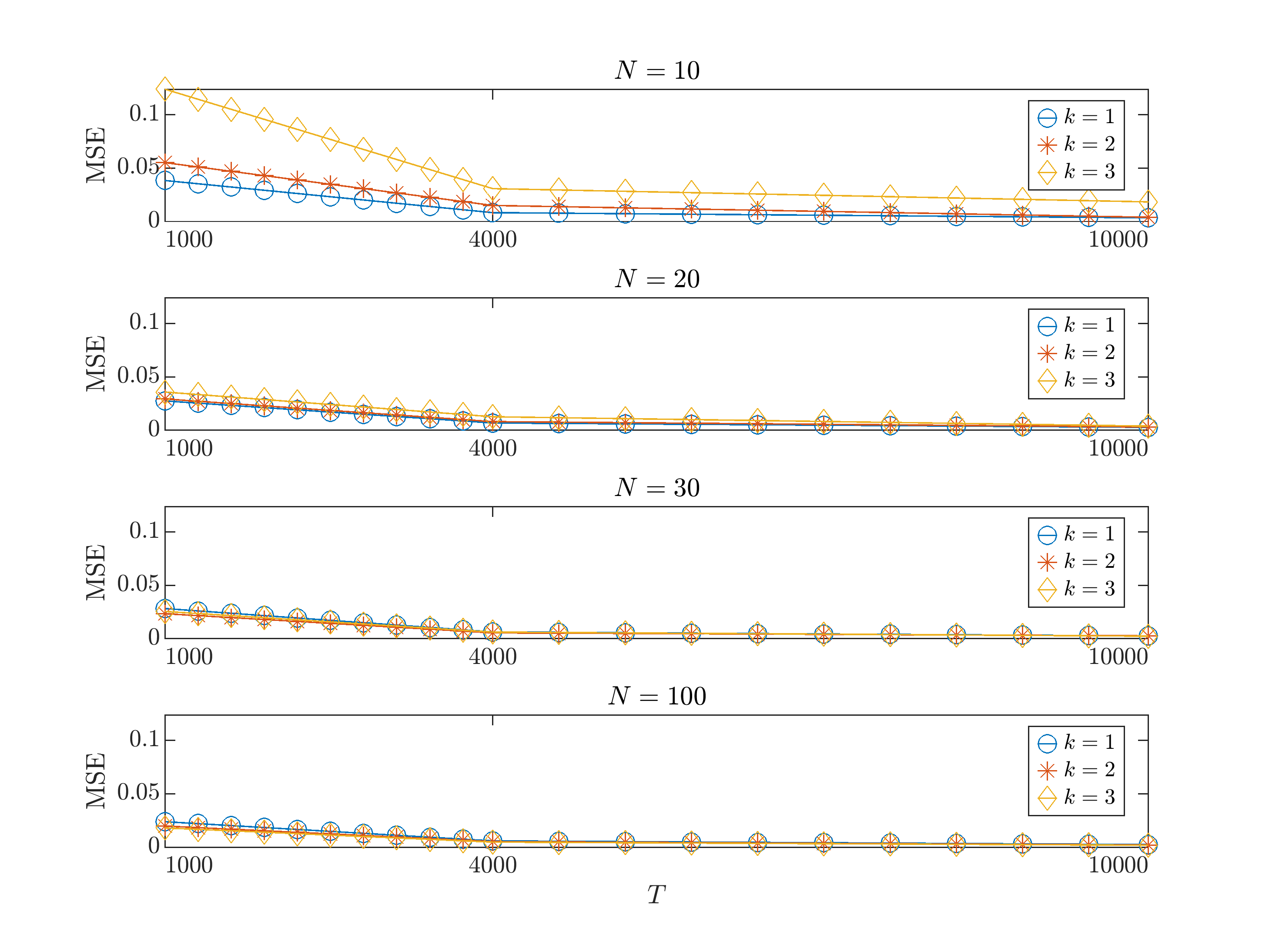}
\caption{MSE of $\hat{\bm\theta}$: $H = 10^{5}/T$ and QML starting values for EMM}\label{MSE_fig_H_adapt}
\end{figure}

Figure \ref{MSE_fig_H_adapt} plots the MSE values for $H = 10^{5}/T$ and QML starting values. The figure plots the MSE against the number of observations $T$ for different numbers of returns series $N$. The numerical MSE results for this case are detailed in table \ref{MSE_tabdet_GARCH_KF_H_adapt} in Appendix \ref{appendix6}. The figure summarises the MSE results averaged over all parameters. In a nutshell: increasing $N$ and $T$ significantly reduces the MSE, while increasing $k$ has a detrimental effect on the MSE, especially if $N$ and $T$ are small.

\subsection{Ordering and computational efficiency}

Due to the identifying assumptions we make for our model, the ordering of the returns in the vector $\bm{y}_{t}$ matters since the first series only loads on the first factor, the second series only loads on the first two factors, and so on. While the factor loadings matrix, the unconditional variance-covariance matrix of the factors and the constant parameter of the factor ARSV models change by changing the order, the other set of parameters should not change and this is confirmed in a small Monte Carlo simulation study where we invert the original ordering of the vector of returns. The results, which are computed with $H=10^5/T$ and QML starting values, can be obtained from the authors upon request.

\begin{figure}[hbt]
	\vspace{1cm}\includegraphics[width = 1\linewidth]{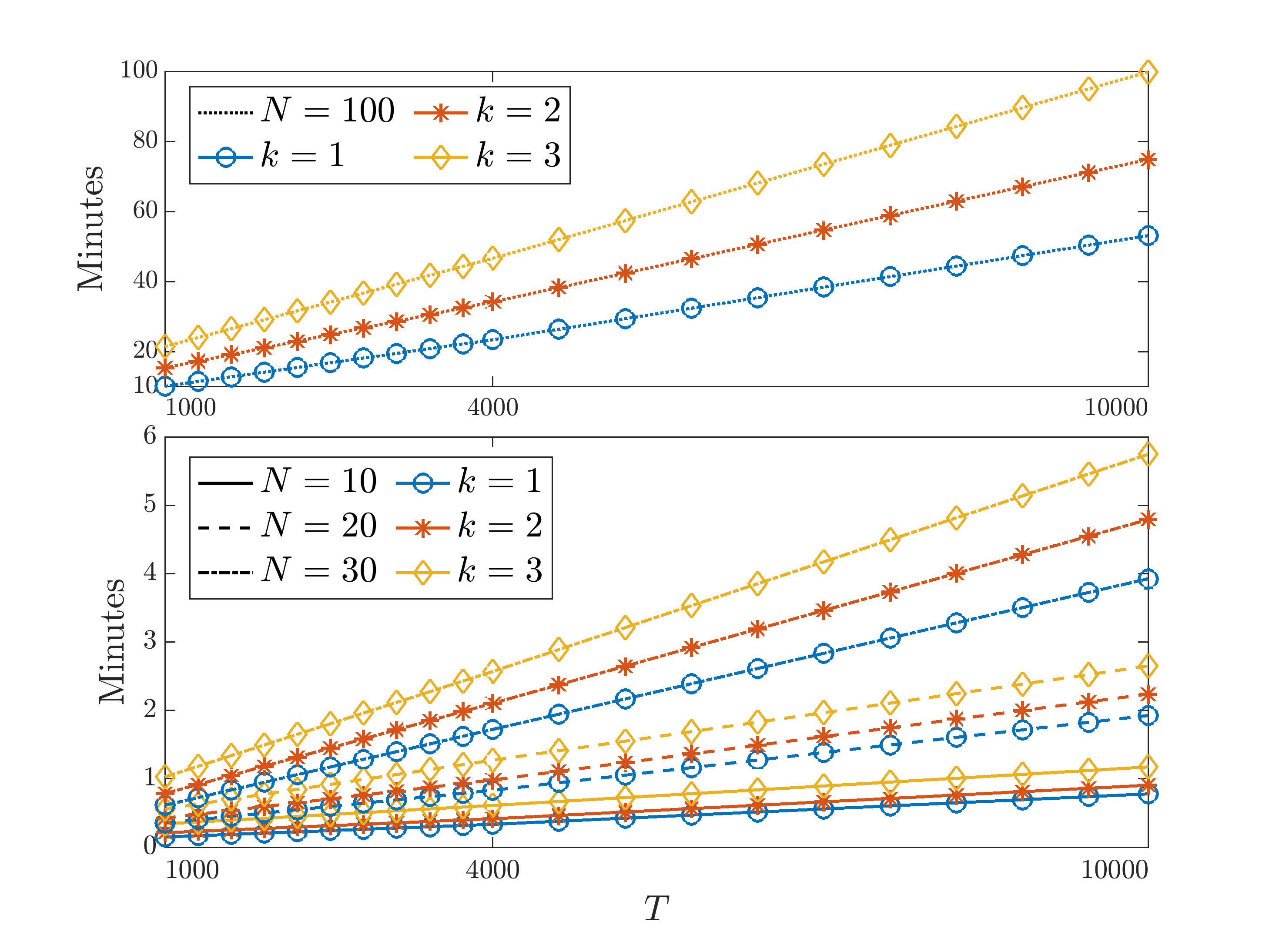}
\caption{Average time per Monte Carlo replication: $H = 10$.}
\label{time_fig_H_10}
\end{figure}

\begin{figure}[hbt]
\vspace{1cm}\includegraphics[width = 1\linewidth]{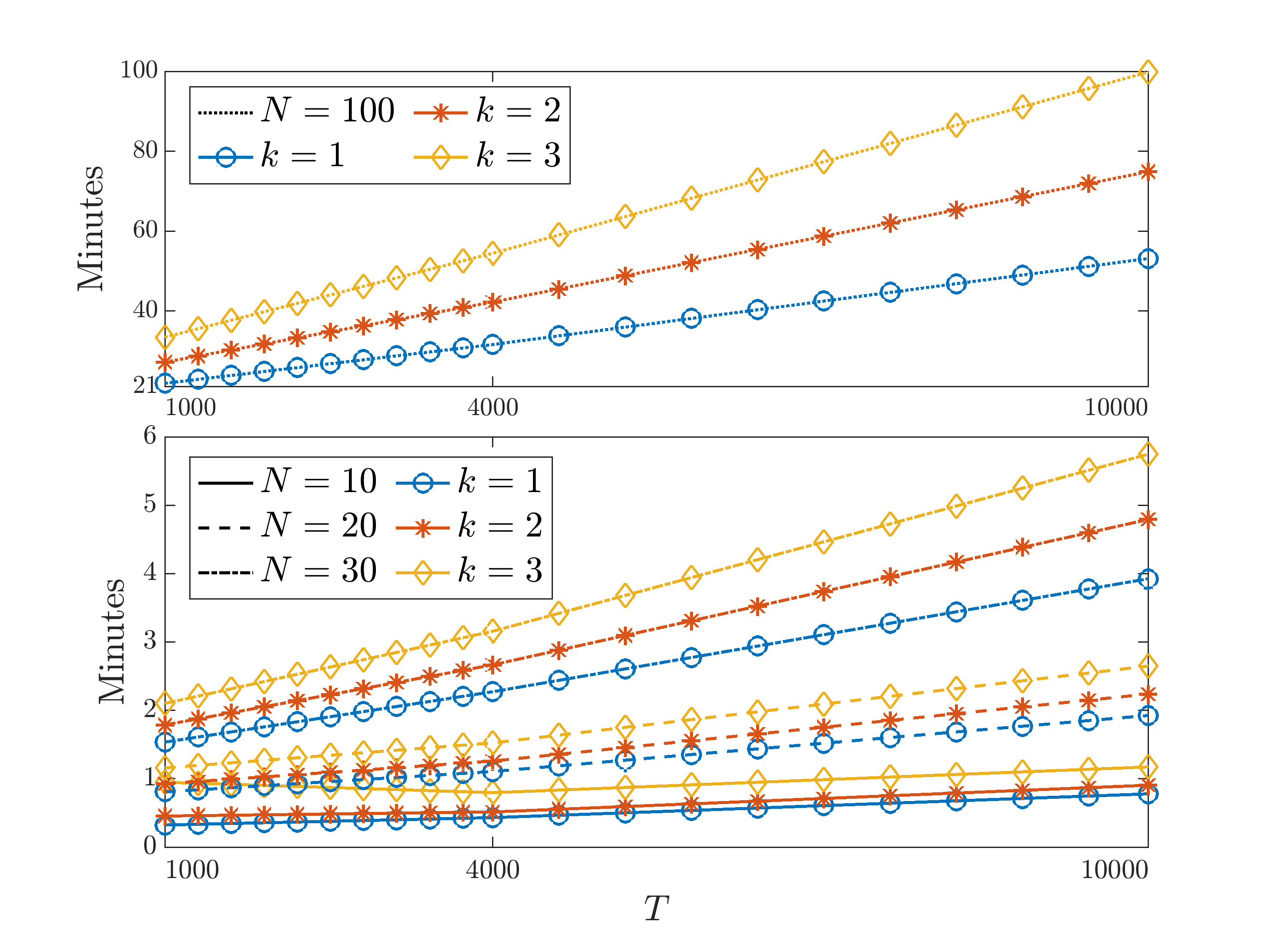}
\caption{Average time per Monte Carlo replication: $H=10^{5}/T$.}
\label{time_fig_H_adapt}
\end{figure}

One very important advantage of our method is its computational efficiency. Figures \ref{time_fig_H_10} and \ref{time_fig_H_adapt} plot the computational time per replication for $H=10$ and $H = 10^{5}/T$, respectively against $T$ for different numbers of factors and return series, however by implementing the QML starting values for the EMM estimation. The upper panel shows the average computational time over the Monte Carlo replications for $N = 100$ return series for one, two, and three factors, while the lower panel depicts smaller dimensions, i.e., $N=10, 20, 30$. 

From the two graphs one may see that the computational time increases linearly in $T$. Moreover, surprisingly the computational time for $H = 10^{5}/T\geq 10$ is only maximally twice the time needed for $H=10$. For instance, for $T=4000$ (i.e., $H = 25$), there are almost no differences in the computational time between the two choices of $H$. Therefore, one can conclude that adapting the choice of $H$ to $T$ is optimal for any empirical application as it increases the efficiency of the estimates, reduces the number of outliers and is computationally feasible.

In order to present results on the computational time of our procedure we focus here on the simulation design of \cite{kastner17} and choose $N = 10$, $k = 2$ and $T = 1000$ as well as their choice of parameters. Our procedure estimates the parameters of the MFSV model under the settings of \cite{kastner17} on average in 13 seconds when we use the QML starting values and set $H=10$. Setting $H = 10^{5}/T$- and with poorer starting values, our procedure takes, on average, 82 seconds (roughly 1.37 minutes). Therefore, we can conclude that our procedure is very effective and computationally feasible in estimating the parameters of MFSV even for very large vectors of returns as also illustrated in the empirical application presented below.

Although it does not make the object of this paper, we also report here some computational time from filtering the unobserved series of the model, namely: $15.7$ minutes using $5000$ particles, $28.43$ minutes using $50000$ particles and $158.2$ minutes using 500000 particles. We obtain these numbers replicating the simulation study of \cite{kastner17}, i.e., the case for $n = 10$, $k = 2$, and $T = 1000$ for $1000$ Monte Carlo replications. We perform the filtering using a straightforward implementation of the Bootstrap Particle Filter of \cite{gordon93}. In comparison, the Bayesian procedure of \cite{kastner17}, which estimates the parameters and filtered values simultaneously, requires 1446.208 minutes (roughly 24 hours) using $550000$ sampling periods, of which we discard the first $50000$ as burn-in periods.\footnote{We replicate their MC study in $R$ using the \textit{factorstochvol} package provided by the authors. We run the code on the bwHPC Cluster. Each job uses 1 core of a Octa-core Intel Xeon E5-2670 (Sandy Bridge) Processors with 2.6GHz), since the package does not allow for parallel execution.}

\section{Empirical Application}\label{real}
In this section, we apply our proposed algorithm to a vector of $N = 148$ log-returns of large-cap stocks from the S\&{P}500 from January 1981 till December 2018 (a total of $T=9584$ observations), ordered by market cap.\footnote{We provide a list of the ticker symbols and company names ordered by the market capitalization of the stocks in Table \ref{symbolnames} in Appendix \ref{appendix7}.} We demean and standardize the returns such that each series has zero sample mean and unit sample variance to facilitate the estimation. Since we have close to $T=10000$ observations, we use the adaptive choice of $H=10^5/T$ and set, thus, it to $H =10$.

As our procedure only regards the estimation of the MFSV model for a prespecified number of factors, we apply the test of \cite{onatski09} to find the number of dynamic factors in our dataset. The test is implemented iteratively with the hypotheses
\begin{equation}
H_{0}: k = k_{1} \qquad \text{vs} \qquad H_{1}: k_{1} < k \leq k_{2},
\end{equation}
where $k_1$ and $k_2$ are pre-specified. We set $k_2=3$ and $k_1=0,1,2$. We run the test using the Matlab code provided by \cite{onatski09} and present the results in Table \ref{onatski_test}. The table clearly shows that we do not reject $H_0$ that $k=1$ at 10\% significance level. Figure \ref{Eig_real} in Appendix \ref{appendix7a} shows the number of principle components of the data set we use in this empirical application. The cutoff between the first and the rest of the principal components is pretty clear, which confirms the results of the test of \cite{onatski09}. That is why we present below first the empirical results for estimating the MFSV model with $k=1$ factors and, afterwards, we discuss the results from estimating the model for $k=2$ and $k=3$. 

\begin{table}[hbt]\centering\caption{P-values for the test of \cite{onatski09} with $k_{1}=0$ and $k_{2}=3$.}\label{onatski_test}
\setlength{\columnsep}{0.2\linewidth}
\begin{tabular}{l | p{0.15\linewidth} | p{0.15\linewidth} | p{0.15\linewidth}}\toprule
$H_{0}$ & {$k=0$ vs. $0<k\leq 3$} & $k=1$ vs. $1<k\leq 3$ & $k=2$ vs. $2<k\leq 3$ \\\midrule
p-value &  0.0150  &   0.1890 &   0.2130\\\bottomrule
\end{tabular}
\end{table}

\subsection{Results for $k=1$}

Figures \ref{real_mui_k1} to \ref{real_setai_k1} in Appendix \ref{appendix8} display the estimated parameters of the ARSV dynamics of the idiosyncratic error term series. While the estimates of the constant parameter seem to vary less among the 148 idiosyncratic noises, the estimates of the autoregressive parameter and of the noise standard deviation display much more variation. Most of the idiosyncratic noises have a very persistent stochastic volatility, with the AR parameter being close to 1, while others display less persistency in their dynamic variation, with the AR parameter being around 0.8. Also the standard deviation of their SVs vary largely among the series from values close to zero to values larger than 0.3. This mirrors the large heterogeneity in the dynamics of the 148 returns considered in the series study

The AR parameter in the SV structure of the factor is estimated to be around 0.9752, while the constant and the standard deviation of the noise are estimated to be -1.5866 and 0.1863, respectively. All parameters of the factor SV are statistically significant at 5\% significance level.

In Table \ref{comptime_1_factor} we show the computational time needed for our estimation procedure to estimate the parameters of the MFSV model with one factor on the vector of 148 daily returns of this empirical application. When unable that the estimation of the second step runs in parallel, the run time is roughly 3,5 hours. Parallelizing the second step on 40 CPU cores, the run time reduces to approximately 10 minutes. Because we have no comparison of our estimation time to any other purely estimation algorithm in the literature, we can only present here the time of convergence of the Bayesian procedure of \cite{kastner17}, which is roughly 59 hours. In order to make the comparison complete, we also present some preliminary results from filtering the latent variables based on our estimates. The classical Bootstrap Particle Filter by \cite{gordon93} we use for the runtime comparison for the $n=10$ and $k=2$ case in the simulations, suffers from the curse of dimensionality and therefore does not yield reliable filtering results. However, in \cite{muecher23}, we introduce the Auxiliary Multiple Particle Filter for the MFSV, to deal with this problem. Applying this filter here with 10240 particles per series (i.e., a total of 1525760 particles) takes 20.5 hours, making the whole procedure roughly 2.9 times faster than the Bayesian procedure. However, the filter uses seven times more particles than the Bayesian procedure uses sampling iterations. \footnote{ We use the \textit{factorstochvol} package provided by the authors for the estimation and, as suggested by them, we use the deep interweaving code. After 20000 burn-in periods we use 200000 draws from the posterior distribution. The authors use 50000 burn-in and 500000 sampling periods in their paper. Our choice of the burn-in and sampling periods is due to RAM and runtime limitations on our computational device.}
\begin{table}[hbt]
\centering\caption{Computational time for MFSV with $k=1$ based on our estimation procedure and based on the Bayesian method of \cite{kastner17}}\label{comptime_1_factor}
\setlength{\columnsep}{0.1\linewidth}
\begin{tabular}{p{6cm} c r @{.} l }
\toprule
& \# CPUs	&\multicolumn{2}{c}{Hours} \\\midrule
\cite{kastner17}&1		&    58&9626  \\\midrule
\multicolumn{4}{c}{Our approach} \\\midrule
{Estimation} 	&1		&     3&5516  \\
				&40 	&	  0&1558  \\\midrule
{Filtration} 	&40 	&	20&4226  \\
\bottomrule
\end{tabular}
\end{table}

\subsection{Results for $k=2$ and $k=3$}
Although as suggested by the test of \cite{onatski09} and PCA, we only need one factor to capture the commonality in the stochastic volatility dynamics of the series, we present here, for completeness, the results from estimating the MFSV model for more factors. We choose $k=3$ to be the maximum, as it has also been a common choice in the existing Bayesian estimation literature\citep{han05,nardari07}. We report the estimation results of our procedure in Appendix \ref{appendix10}, namely in Table \ref{real_factors} for estimating the SV parameters of the factors and in figures \ref{real_mui} to \ref{real_setai}, for estimating the SV parameters of the idiosyncratic error series. 

The first finding of applying our estimation procedure to MFSV models with more factors that the "true" ones is that it detects this mismatch by delivering not significant estimates or anomalies in the estimates, as reported in Table \ref{real_factors}: while the pattern of the estimates of the SV model for the idiosyncratic noises is similar to the ones of $k=1$ and the AR parameters of the SV model of the factors are all very close to 1, which indicate high persistency in the volatility dynamics of the series, similar to the case $k=1$, most of the SV parameters of the second and third factor are estimated to be not significant and some of them have unusual high standard errors. This might be due to the fact that the accurate ML estimation of a static factor model in the first step requires some degree of variation in the factors (\citealp{bai12}; \citealp{bai16}). Suppose this is not the case, as it happens when estimating more factors than necessary, which in fact, should have variation zero. In that case, the rank condition on the factor loadings matrix is violated \citep{anderson56}, and the distribution of the static factors is not identified \citep{cox17}. In other words, estimating factors with almost zero variance may indicate that the model is not correctly specified and that there are fewer factors describing the true structure. Therefore, the empirical results for the case of $k=2$ and $k=3$ should be handled with care as they refer to misspecified MFSV models. 

To bring further evidence in this direction, we run a small simulation study where we simulate $N=10$ and $N=100$ return series with $T = 4000$ observations and $k = 2$ factors. Figures \ref{Eig_N10} and \ref{Eig_N100} in Appendix \ref{appendix10} show the first 10 principal components of the $R=1000$ simulated data sets for $N = 10$ and $N = 100$ respectively. The plots clearly indicate the number of factors to be equal to 2, which is precisely the number of factor we simulate. On this simulated series, we apply our estimation procedure by choosing $k = 1,2,3$ factors. We find that the Fisher information matrix of the auxiliary model for $k=3$ is not of full rank with serious effects on the standard errors and the inference based on them. In fact for $k=1$ and $k = 2$ the estimation works smoothly, while for $k=3$, the first step estimation reveals zero variance of the third factor, as depicted in Figure \ref{gamma_N10} in Appendix \ref{appendix10} for $N=10$ and in Figure \ref{gamma_N100} for $N=100$.

The fact that for $k=2$ and $k=3$ we deal with misspecified MFSV models for our real dataset is also detectable when running the Bayesian procedure of \cite{kastner17}, which in fact, sets the constant in the factor ARSV model to 0 and instead specifies a scaling parameter for the variance of the factor. The variance of the factor is multiplied by this scaling parameter \citep{broto04}. In the specific context of the MFSV model, the scaling parameters are multiplied then with the respective column of the loadings matrix, allowing for diagonal elements that are not restricted to 1. Figures \ref{kastner_trace} and \ref{kastner_dens} in Appendix \ref{appendix10} show that, for $k=3$ the scaling of the second factor is ill-behaved. Specifically, we see in Figure \ref{kastner_trace} that the Markov Chain does not converge for the scaling factor and Figure \ref{kastner_dens} shows that the estimated marginal density of the scaling factor is bi-modal, with a lot of probability mass around zero. Thus, the Bayesian procedure also faces "problems" when the number of factors becomes larger than the true one.


\section{Conclusion}\label{conclusion}
This paper proposes a two step procedure to estimate multivariate factor stochastic volatility models with exact factor structure and stochastic volatility dynamics for both the factors and idiosyncratic noises. In the first step we use use the convergence results of \cite{bai12} and \cite{bai16} to estimate the factor loadings and the variances of the idiosyncratic errors and of the factors by means of ML applied to a static factor model. In the second step we estimate the parameters of the stochastic volatility of the factors and of the idiosyncratic noises equation-by-equation by using the EMM procedure of \cite{bansal94} and \cite{gallant96} and univariate GARCHs as auxiliary models. 

Our procedure is the first in the literature to implement "frequentist" techniques to estimate such a model. Due to the two-step procedure and equation-by-equation estimation in the second step, the procedure we propose is computationally very fast and can easily be applied to very large dimensions of vectors of returns and number of factors. The computational efficiency can be further increased by choosing appropriate starting values and parallelizing the EMM estimation in the second step. 

In a comprehensive simulation exercise we show that our estimation provides accurate estimates of the parameters as well as how to choose the optimal number of simulations in the EMM procedure in order to increase the efficiency of the estimation at minimal computational costs. In the empirical application to a large vector of 148 returns we provide further evidence of the efficacy of our method. While in the simulation exercise we can estimate the model to a vector of 100 returns in a matter of seconds, in the empirical application the estimation converges in about 10 minutes. 


\section*{Acknowledgments}
Cristian M\"ucher acknowledges financial support from the Graduate School of Decision Sciences (GSDS), University of Konstanz, Germany and from the German federal state of Baden-W{\"u}rttemberg through a Landesgraduiertenstipendium. Roxana Halbleib acknowledges financial support from the German Science Foundation through the project HA 8672/1. We want to thank Robin Braun, Ralf Br\"uggemann, Maurizio Daniele, Giampiero Gallo, Lyudmila Grigoryeva, Julie Schnaitmann, Winfried Pohlmeier, the participants of the SoFiE Conference 2019, QFFE International Conference 2019, CFE 2018 and Econometrics Colloquium at the University of Konstanz, for helpful comments. All remaining errors are ours. We acknowledge computational support by the state of Baden-W\"urttemberg through bwHPC.

\newpage
\bibliographystyle{apalike}
\bibliography{ref_com_chris}
\newpage
\appendix 
\section*{Appendix}
\section{Estimation of the Factor Model Parameters by ML}
\label{appendix1}

 \cite{bai12} apply the EM algorithm on a slightly different static factor model than ours defined in \eqref{eq1} to \eqref{eq3}; i.e., one that fulfills the identification restrictions of the Principle Components Analysis (PCA):
 \begin{align}\label{eq12}
\bm{y}_{t} &= \underline{B}^{*}\bm{g}^*_{t} + \bm{e}_{t} \\
\label{eq22}\bm{e}_{t} &\overset{i.i.d}{\sim} \mathcal{N}\left(\bm{0}_{N\times 1}, \Sigma^{*}\right),\\
\label{eq32} \bm{g}^*_{t} & \overset{i.i.d}{\sim}\mathcal{N}\left(\bm{0}_{k\times 1},I_{k}\right),
\end{align}
where $I_{k}$ is a $k$ dimensional identity matrix. Thus, in the model (\ref{eq12})-(\ref{eq32}), the identification assumptions are that the diagonal elements of $(1/N){\underline{B}^{*}}\T(\Sigma^{*})^{-1}\underline{B}^{*}$ are distinct and positive and are arranged in decreasing order. $\underline{B}^{*}$ is identified up to a column sign change \citep{bai12}. 

One may see that while Equation (\ref{eq22}) is identical to Equation (\ref{eq2}), meaning that in the PCA identification framework the error inherits the structure of the one in the exact static factor model, the loadings and factors/components differ slightly due to other identification restrictions. The EM algorithm we use in this paper is applied to the model (\ref{eq12})-(\ref{eq32}), however, we ex-post use the rotation procedure described by \cite{bai12} to recover the parameters of our static factor model defined in equations \eqref{eq1} to \eqref{eq3}, as described below.

For convenience, we present here the estimation based on the negative log-likelihood of the model specified in equations \eqref{eq12}, \eqref{eq22} and \eqref{eq32} that is proportional to 
\begin{equation}
F^{a,1}(\bm{y}_{t}, {\underline{B}}^{*}, \Sigma^{*}) = \frac{1}{N}\underbrace{\ln{\left\{\det\left({\underline{B}}^{*}{{\underline{B}}^{*}}\T + \Sigma^{*}\right)\right\}}}_{M^{*}} + \frac{1}{N}\underbrace{\tr \left[\left(\underline{B}^{*}{\underline{B}^{*}}\T + \Sigma^{*}\right)^{-1}A\right]}_{C^{*}},\label{Likelihood}
\end{equation}
with $A$ being the empirical covariance matrix of the data. Thus, the specification in Equation (\ref{Likelihood}) is a combination of a concave part (denoted by $M^{*}$) and a convex part (denoted by $C^{*}$), which usually leads to many local minima in the optimization \citep{bai12, bien11}. We describe below the steps of implementing the EM algorithm of \cite{bai12} to circumvent this problem, however not directly to the proportional likelihood in Equation \eqref{Likelihood}, but to the "majorized" negative log-likelihood proposed by \citet{bien11} that approximates the concave part of \eqref{Likelihood}, $M^*$ by its tangent hyperplane at each iteration $l$:
\begin{align*}
\tilde{F}^{a,1}(\bm{y}_{t}, {\underline{B}}^{*}, \Sigma^{*})= &\frac{1}{N}\ln{\left|\det\left(\widehat{\underline{B}^{*}_{\ell}}{\widehat{\underline{B}^{*}_{\ell}}}\T + \widehat{\Sigma^{*}_{\ell}}\right)\right|}+\frac{1}{N} \tr\left[2{\widehat{\underline{B}^{*}_{\ell}}}\T\left(\widehat{\underline{B}^{*}_{\ell}}{\widehat{\underline{B}^{*}_{\ell}}}\T + \widehat{\Sigma^{*}_{\ell}}\right)^{-1}(\underline{B}^{*}-\widehat{\underline{B}^{*}_{\ell}})\right] \\
&+\frac{1}{N} \tr\left[(\underline{B}^{*}{\underline{B}^{*}}\T + \widehat{\Sigma^{*}_{\ell}})^{-1}A\right],\numberthis\label{majorizedlike}
\end{align*}
where $\widehat{\underline{B}^{*}_{\ell}}$ and $\widehat{\Sigma^{*}_{\ell}}$ are the estimators at the $\ell$-th iteration of the EM algorithm obtained at S2, described below,
yielding, thus, a fully convex optimization problem. After the algorithm converges, we follow \cite{bai12} and rotate $\widehat{\underline{B}^{*}}$ to obtain $\widehat{B^{*}}$ and $\widehat{\Gamma^{*}}$, which correspond to the static factor of our interest defined in equations \eqref{eq1} to \eqref{eq3}. 
The EM algorithm is implemented through the following steps \citep{bai12}:

\begin{itemize}
	\item[\textit{S1:}] Get the initial values $\widehat{\underline{B}^{*}_{0}}$ and $\widehat{\Sigma^{*}_{0}}$ from PCA.
	
	\item[\textit{S2:}] Update $\widehat{\underline{B}^{*}_{\ell}}$ by 
\begin{equation}\widehat{\underline{B}^{*}_{\ell+1}} = \widehat{\underline{B}^{*}_{\ell}}-d\widehat{D}_{\ell},\end{equation}
	with 
	\begin{equation}\widehat{D}_{\ell} = 2\left[\left(\widehat{\underline{B}^{*}_{\ell}}\widehat{{\underline{B}^{*}_{\ell}}}\T + \Sigma^{*}_{\ell}\right)^{-1}-\left(\widehat{\underline{B}^{*}_{\ell}}\widehat{{\underline{B}^{*}_{\ell}}}\T + \Sigma^{*}_{\ell}\right)^{-1}A\left(\widehat{\underline{B}^{*}_{\ell}}\widehat{{\underline{B}^{*}_{\ell}}}\T + \Sigma^{*}_{\ell}\right)^{-1}\right]\widehat{{\underline{B}^{*}_{\ell}}}\end{equation}
being the partial derivative of \eqref{majorizedlike} with respect to $\underline{B}^{*}$ and $d$ being the step size of the Gradient Descent Algorithm \citep{bien11}.
	\item[\textit{S3:}] Update $\widehat{\Sigma^{*}_{\ell}}$ by 
\begin{equation}\widehat{\Sigma^{*}_{\ell+1}} = \diag\left(A-\widehat{\underline{B}^{*}}_{\ell+1}{\widehat{\underline{B}^{*}}}_{\ell}\T\left(\widehat{\underline{B}^{*}}_{\ell}\widehat{{\underline{B}^{*}}}_{\ell}\T + \Sigma^{*}_{\ell}\right)^{-1}A\right).\end{equation}
	\item[\textit{S4:}] Repeat steps 2 and 3 until $\norm{\widehat{\underline{B}^{*}}_{\ell+1}-\widehat{\underline{B}^{*}_{\ell}}}_{F}$ and $\norm{\widehat{\Sigma^{*}_{\ell+1}} - \widehat{\Sigma^{*}_{\ell}}}_{F}$, where $\norm{\cdot}_{F}$ denotes the Frobenius norm, are smaller than some threshold. Denote the values of $\widehat{\underline{B}^{*}_{\ell}}$ and $\widehat{\Sigma^{*}_{\ell}}$ from the final iteration (up to convergence) by $\widehat{\underline{B}}^{*}$ and $\widehat{\Sigma^{*}}$.
	\item[\textit{S5:}] Rotate $\widehat{\underline{B}}^{*}$ according to:
	\begin{equation}\widehat{B^{*}} = \widehat{\underline{B}^{*}}\mathcal{Q}\mathcal{W}^{-1},\end{equation}
	with $\mathcal{Q}$ stemming from the QR-decomposition of the transpose of the first $k\times k$ block of $\widehat{\underline{B}}^{*}$ and $\mathcal{W}= \diag( \widehat{\underline{B}^{*}}\mathcal{Q})$.
	\item[\textit{S6:}] Compute the estimator for the unconditional factor variances:
\begin{equation}\widehat{\Gamma^{*}} = \mathcal{WW}\T.\end{equation}
\end{itemize}

$\widehat{B^{*}}$, $\widehat{\Sigma^{*}}$ and $\widehat{\Gamma^{*}}$ obtained in $S5$ and $S6$ by rotating $\widehat{\underline{B}}^{*}$ obtained in $S4$ fulfill the identification conditions of our static factor model defined in equations \eqref{eq1} to \eqref{eq3} \citep{bai12}.

\section{The EMM Estimation of the ARSV parameters}
\label{appendix2}
The ARSV parameters $\bm{\theta}_{2} = \left(\varphi_{1},\hdots,\varphi_{N+k},\sigma_{\eta,1}, \hdots,\sigma_{\eta,N+k}\right)'$ are estimated in pairs, i.e., $\bm{\theta}_{2,m} = (\varphi_{m},\sigma_{\eta,m})'$ by making use of the auxiliary GARCH(1,1) estimation applied equation-by-equation to each element of $\hat{\bm{x}}_{m} = (\hat{x}_{1,m},\hdots,\hat{x}_{T,m})\T$ with $m=1,\ldots, N+k$. Thus, the auxiliary log-likelihood is proportional to:
		\begin{equation}
\mathcal{L}^{a,2}_m(\hat{\bm{x}}_{m};\boldsymbol{\beta}_{2,m}) = -\frac{1}{T}\sum\limits_{t = 1}^{T}\left(\ln{(\delta_{t,m}^{2}(\boldsymbol{\beta}_{2,m}))} +\frac{\hat{x}_{t,m}^{2}}{\delta_{t,m}^{2}(\boldsymbol{\beta}_{2,m})}\right),\label{garchlik} 
\end{equation}
	where $\delta_{t,m}^{2}$ is defined in Equation (\ref{garch}) and it is a function of $\boldsymbol{\beta}_{2,m}$.
		The single contributions, i.e., at time $t$, to the auxiliary model score vector, are obtained by taking the derivative of the log-likelihood function in (\ref{garchlik}) with respect to the parameter vector $\boldsymbol{\beta}_{2,m}$ at time $t$
\begin{eqnarray}q_{t,m}(\hat{x}_{t,m},\bm{\beta}_{2,m}) &=& \frac{\partial \mathcal{L}^{a,2}_m(\hat{\bm{x}}_{m};\boldsymbol{\beta}_{2,m})}{\partial \bm{\beta}_{2,m}}.\\
&=& \frac{1}{\delta_{t,m}^{2}(\boldsymbol{\beta}_{2,m})}\left(\frac{\partial \delta_{t,m}^{2}(\boldsymbol{\beta}_{2,m})}{\partial \boldsymbol{\beta}_{2,m}}\right)\left(\frac{\hat{x}_{t,m}^{2}}{\delta_{t,m}^{2}(\boldsymbol{\beta}_{2,m})} - 1\right),\label{dqgarch}
\end{eqnarray}
where \begin{equation}
\frac{\partial \delta_{t,m}^{2}(\boldsymbol{\beta}_{2,m})}{\partial \boldsymbol{\beta}_{2,m}} = \left(\begin{array}{c} \hat{x}_{t-1,m}^{2}\\ \delta_{t-1,m}^{2}(\boldsymbol{\beta}_{2,m})\end{array}\right) - \hat\psi_{m} + \alpha_{2,m} \frac{\partial \delta^{2}_{t-1,m}(\boldsymbol{\beta}_{2,m})}{\partial \boldsymbol{\beta}_{2,m}}\label{dsigmagarch}.
\end{equation}
The score vector is then given by taking the mean over the single contributions in Equation (\ref{dsigmagarch}):
\begin{equation}\mathcal{Q}_{m}(\hat{\bm{x}}_{m};\bm{\beta}_{2,m}) = \frac{1}{T}\sum_{t=1}^{T}q_{t,m}(\hat{x}_{t,m},\bm{\beta}_{2,m}).\end{equation}

As described below, the EMM estimation for each of the pair of parameters $\bm{\theta}_{2,m} = (\varphi_{m},\sigma_{\eta,m})'$ requires $H$ number of simulated series $\bm{y}_{t}$. Section \ref{estimation} provides a detailed discussion on how to choose $H$ for our study. 

Thus, the EMM estimation of the ARSV parameters can be parallelized over $m=1,\ldots, N+k$ and consists of the following steps:
\begin{itemize}
	\item[\textit{A0:}] Set $m=1$ and choose the $m$-th time series of observations from \\
	$\hat{\bm{x}}_{t} = (\hat{e}_{t,1},\hdots,\hat{e}_{t,N}, \hat{g}_{t,1},\hdots,\hat{g}_{t,k})'$.
		\item[\textit{A1:}] Get the PML estimate $\hat{\bm{\beta}}_{2,m}$ of the GARCH(1,1) model based on $\mathcal{L}^{a,2}_m(\hat{\bm{x}}_{m};\boldsymbol{\beta}_{2,m})$.
						\item[\textit{A2:}] Evaluate the auxiliary model score at $\hat{\bm{x}}_{m}$ and $\hat{\bm{\beta}}_{2,m}$, $\mathcal{Q}_{m}(\hat{\bm{x}}_{m};\hat{\bm{\beta}}_{2,m})$.
		\item[\textit{A3:}] Choose a vector of initial values $\bm{\theta}_{2}$, i.e., $\bm{\theta}_{2}^{0}$.
		\item[\textit{A4:}] Simulate $T\times H$ observations $\tilde{\bm{y}}_{t}(\hat{\bm{\theta}}_{1},\bm{\theta}_{2}^{0})$ from the MFSV data generating process defined in Section \ref{MFSV} by using $\hat{\bm{\theta}}_{1}$ obtained at the first step and $\bm{\theta}_{2}^{0}$.
		\item[\textit{A5:}] Compute $\tilde{\bm{x}}(\hat{\bm{\theta}}_{1},\bm{\theta}_{2}^{0})$ from equations \eqref{projection} and \eqref{residuals}, by replacing $\bm{y}_{t}$ with its' simulated counterpart $\tilde{\bm{y}}_{t}(\hat{\bm{\theta}}_{1},\bm{\theta}_{2}^{0})$ and by fixing $\Pi^*$ in Equation (\ref{projection}) for all iterations and all $m$'s to the estimated values from the first estimation step described in Section \ref{part1}.
		\item[\textit{A6:}] Evaluate the auxiliary model score at the $m$-th element of $\tilde{\bm{x}}(\hat{\bm{\theta}}_{1},\bm{\theta}_{2}^{0})$, $\tilde{\bm{x}}_{m}(\hat{\bm{\theta}}_{1},\bm{\theta}_{2}^{0})$ and at $\hat{\bm{\beta}}_{2,m}$: $\mathcal{Q}_{m}(\tilde{\bm{x}}_{m}(\hat{\bm{\theta}}_{1},\bm{\theta}_{2}^{0});\hat{\bm{\beta}}_{2,m})$.
		\item[\textit{A7:}] Update the $m$-th element of $\bm{\theta}_{2}^{0}$, i.e., $\bm{\theta}_{2,m}^{0}$ to $\bm{\theta}_{2,m}^{1}$, by keeping the rest of the $m-1$ elements of $\bm{\theta}_{2}^{0}$ at the initial values set at step \textit{A3}. Denote the updated $\bm{\theta}_{2}$ vector by $\bm{\theta}_{2}^{1_m,0}$ and repeat steps \textit{A4} to \textit{A6} until the weighted squared distance: 
		\begin{align*}
&\left(\mathcal{Q}_{m}(\tilde{\bm{x}}_{m}(\hat{\bm{\theta}}_{1},\bm{\theta}_{2}^{1_m,0});\hat{\bm{\beta}}_{2,m}) - \mathcal{Q}_{m}(\hat{\bm{x}}_{m};\hat{\bm{\beta}}_{2,m})\right)\T\Omega \\
&\left(\mathcal{Q}_{m}(\tilde{\bm{x}}_{m}(\hat{\bm{\theta}}_{1}\bm{\theta}_{2}^{1_m,0});\hat{\bm{\beta}}_{2,m}) -\mathcal{Q}_{m}(\hat{\bm{x}}_{m};\hat{\bm{\beta}}_{2,m})\right)\numberthis,		\end{align*} 
where $\Omega$ is a positive definite weighting matrix, is minimized. The $m$th element of the last updated $\bm{\theta}_{2}^{j_m,0}$ that minimizes the distance is the EMM estimator of $\bm{\theta}_{2,m}$, i.e., $\hat{\bm{\theta}}_{2,m}$.
 \item[\textit{A8:}] Set $m=m+1$ and repeat steps \textit{A1-A7} until $m=N+k$.\footnote{We also implement an iterative version of the estimation procedure, where we use the estimated values at step \textit{A8} as new initial values at step \textit{A3}. We then iterate these steps until convergence. However, we find that the estimation results do not change up to three significant digits after the comma.}
\end{itemize}
In our case, the parameter dimension of the auxiliary model is the same as the one of MFSV, i.e., $\dim{\bm{\theta}} = \dim{\bm{\beta}}$. Particularly, we have that $\dim{\bm{\theta_{2,m}}} = \dim{\bm{\beta_{2,m}}}$. Thus, the choice of the weighting matrix $\Omega$ at step \textit{A7} can be arbitrary and we set it equal to the identity matrix. In this case of the exact identified model, the minimization of the weighted squared distance boils down to the minimization of the squared distance. In a matter of fact, in our estimation, we find $\hat{\bm{\theta}}_{2,m}$ that sets $\mathcal{Q}_{m}(\tilde{\bm{x}}_{m}(\hat{\bm{\theta}}_{1},\bm{\theta}_{2}^{j_m,0});\hat{\bm{\beta}}_{2,m}) - \mathcal{Q}_{m}(\hat{\bm{x}}_{m};\hat{\bm{\beta}}_{2,m})=\bm{0}_{2\times 1}$. However, sometimes upon convergence, the $\hat{\bm{\theta}}_{2,m}$ do not fulfill the parameter constraints we define in Section \ref{MFSV} and therefore, in these cases, we optimize the squared distance to get the EMM estimators.

In the steps \textit{A4} to \textit{A7}, we use Equation \eqref{uncondvarARSV} to compute and update the values of the constant of the ARSV process $\mu^{j}_{m}$ from the updated $m$-th element of $\bm\theta_{2}^{j}$, i.e., $\bm\theta_{2,m}^{j}$ and from $\hat{\psi}_{m}$, that is computed in the first estimation step described in Section \ref{part1}.

\section{Closed-form gradient of the static factor model}\label{appendix3}
Consider the static factor model of the first part of the estimation approach:
\begin{align}
\bm{y}_{t} &= B^{*}\bm{g}_{t} + \bm{e}_{t} \label{stat1}\\
\bm{e}_{t} &\sim \mathcal{N}\left(\bm{0}_{N\times 1}, \Sigma^{*}\right)\\
 \bm{g}_{t} & \sim\mathcal{N}\left(\bm{0}_{k\times 1},\Gamma^{*}\right) \label{stat3},
\end{align}
where $\Sigma^{*}$ and $\Gamma^{*}$ are diagonal matrices and $B^{*}$ is restricted such that $b^{*}_{jj} = 1 \text{ and } b^{*}_{ij} = 0 \, \forall j > i, \hspace{3mm} j = 1,\hdots,k, \hspace{3mm} i=j,\ldots,N$. Let $ C= B^{*}\Gamma^{*}{B^{*}}\T + \Sigma^{*}$, then the log-likelihood of the model defined above at time $t$ is proportional to:\footnote{Please note that the static factor in (\ref{stat1})- (\ref{stat3}) differ from the static factor in (\ref{eq12})-(\ref{eq32}) in terms of the identification constraints, as explained in the main text of the paper in Section \ref{estimation}.}
\begin{equation}
F^{a,1,*}(\boldsymbol{y}_{t};\boldsymbol{\beta}_{1}) = -\frac{1}{N}\ln{(\det{C})}-\frac{1}{N}\boldsymbol{y}_{t}\T C^{-1}\boldsymbol{y}_{t}.\label{factorll} 
\end{equation}
Remember that:
\begin{equation}\bm{\beta}_{1} = (b_{21}^{*},\hdots,b_{Nk}^{*},{\sigma_{1}^{*}}^{2},\hdots,{\sigma_{N}^{*}}^{2},{\gamma_{1}^{*}}^{2},\hdots,{\gamma_{k}^{*}}^{2}).\end{equation}
The derivative of Equation \eqref{factorll} at time $t$ with respect to the $p$-th element of of $\bm{\beta}_{1}$, i.e., $\beta_{1,p}$
is given by
\begin{equation}
\frac{\partial F^{a,1,*}(\boldsymbol{y}_{t};\boldsymbol{\beta}_{1})}{\partial\beta_{1,p}} = -\tr{\left(C^{-1} \frac{\partial C}{\partial \beta_{1,p}}\right)} + \tr{\left(C^{-1}\frac{\partial C}{\partial\beta_{1,p}}C^{-1}\frac{\boldsymbol{y}_{t}\T \boldsymbol{y}_{t}}{N}\right)}.\label{stat_fact_score}
\end{equation}
Taking the sum over the $t = 1,\hdots,T$ elements of \eqref{stat_fact_score} yields the score of the static factor model. For each $p$,
$\frac{\partial C}{\partial\beta_{1,p}}$ is an $N\times N$ matrix of derivatives that are defined below, depending on what $\beta_{1,p}$ represents: loadings (\textit{Case 1}), unconditional variances of the factors (\textit{Case 2}) or unconditional variances of the idiosyncratic error terms (\textit{Case 3}):

\textit{Case 1:} $\beta_{1,p} = b_{ij}^{*}$, $\forall j = 1,\hdots,k$ and $i = j,\hdots,N$.\\

Denote by $B^{*}_{j}$ the $j$-th column of $B^{*}$ and let 
\begin{equation}a_{j}= \gamma_{j}^{*}B^{*}_{j}
\end{equation}
be a vector of dimension $N\times 1$. Then $\frac{\partial C}{\partial b^{*}_{ij}}$ is an $N\times N$ matrix built as the sum of two $N\times N$ matrixes, in such a way that the first matrix has all elements zeros except for the $i-$th column which is given by the vector $a_j$ and the second matrix has all elements zeros expect for the $i$-th row that is given by $a_j\T$.\\
 Below we provide an illustration for the matrix $\frac{\partial C}{\partial b^{*}_{21}}$:
\begin{equation}\frac{\partial C}{\partial b^{*}_{21}}=\left(\begin{array}{ccccc}
0 		& a_{1} 	& 0 	& \hdots & 0 \\
0	& a_{2} 	& 0 & \hdots & 0\\
0 		& a_{3} 	& 0 	& \hdots & 0\\
\vdots	& \vdots	& \vdots & 		& \vdots\\
0		& a_{N}		&  0 	& \hdots & 0\\
\end{array}\right)+\left(\begin{array}{ccccc}
0 		& 0	& 0 	& \hdots & 0 \\
a_{1} 	& a_{2} 	& a_{3} & \hdots & a_{N}\\
0 		& 0 	& 0 	& \hdots & 0\\
\vdots	& \vdots	& \vdots & 		& \vdots\\
0		& 0		&  0 	& \hdots & 0\\
\end{array}\right)= \left(\begin{array}{ccccc}
0 		& a_{1} 	& 0 	& \hdots & 0 \\
a_{1} 	& 2a_{2} 	& a_{3} & \hdots & a_{N}\\
0 		& a_{3} 	& 0 	& \hdots & 0\\
\vdots	& \vdots	& \vdots & 		& \vdots\\
0		& a_{N}		&  0 	& \hdots & 0\\
\end{array}\right)
\end{equation}

\textit{Case 2:} $\beta_{1,p} = {\sigma^{*}_{i}}^{2}$, $\forall i = 1,\hdots,N$. \\

$\frac{\partial C}{\partial {\sigma^{*}_{i}}^{2}}$ is an $N\times N$ matrix of zeros, except for the $i$-th diagonal element, which is equal to 1.

\textit{Case 3:} $\beta_{1,p} = {\gamma^{*}_{j}}^{2}$, $\forall j = 1,\hdots,k$.\\

\begin{equation}\frac{\partial C}{\partial {\gamma^{*}_{j}}^{2}}= B^{*}A_j{B^{*}}\T,
\end{equation}
where $A_j$ is a $k\times k$ matrix of zeros with the $j-$th diagonal element equal to 1. 

\FloatBarrier
\clearpage
\newpage
\section{Outliers}\label{appendix4}
\foreach \sn\si in {std/starting values 20\% from the true values,KF/QML starting values} {
	\foreach \hn\hi in {10/10,adapt/10^{5}\slash T} {
		\begin{table}[ht!]\centering{\scriptsize
			\caption{Percentage of outliers for $H = \hi$ and \si}
			\label{GARCH_NC_\sn _H_\hn }
			\ExecuteMetaData[Proc/out/H_\hn /mean/GARCH_\sn _NC_mean_H_\hn .tex]{mytag}
		}\end{table}
}
}
\FloatBarrier
\clearpage
\newpage

\section{Variance Covariance Estimation Results}\label{appendix5}
\foreach \sn\si in {std/starting values 20\% from the true values,KF/QML starting values} {
	\foreach \hn\hi in {10/10,adapt/10^{5}\slash T} {
		\begin{table}[ht!]{\scriptsize
			\caption{Average ratios between the Monte Carlo standard deviation and the asymptotic standard error of the corresponding parameter vectors: $H = \hi$ and \si}
			\label{GARCH_rat_\sn _H_\hn }
			\ExecuteMetaData[Proc/out/H_\hn /mean/GARCH_\sn _rat_RES_mean_H_\hn .tex]{mytag}\\[8pt]}
\scriptsize{Note: The ratios for $B,\Sigma, \Gamma$ are computed by stacking the elements of each matrix in a vector. $\mu_{\varepsilon}\equiv(\mu_{1},\hdots,\mu_{N})\T$, $\mu_{f}\equiv(\mu_{N+1},\mu_{N+2},\mu_{N+3})\T$, $\varphi_{\varepsilon}\equiv(\varphi_1,\hdots,\varphi_N)' $, $\varphi_{f}\equiv(\varphi_{N+1}, \varphi_{N+2}, \varphi_{N+3})'$, $\sigma_{\eta,\varepsilon}\equiv(\sigma_{\eta,1},\hdots,\sigma_{\eta,N})' $ and $\sigma_{\eta,f}\equiv({\sigma}_{\eta,N+1}, {\sigma}_{\eta,N+2},{\sigma}_{\eta,N+3})'$.}
		\end{table}	
}
}
\FloatBarrier

\section{MSE Results}\label{appendix6}
\foreach \sn\si in {std/starting values 20\% from the true values,KF/QML starting values} {
	\foreach \hn\hi in {10/10,adapt/10^{5}\slash T} {
		\begin{table}[ht!]{\scriptsize
			\caption{MSE of the corresponding estimated MFSV parameter vectors: $H = \hi$ and \si}
			\label{MSE_tabdet_GARCH_\sn _H_\hn }
			\ExecuteMetaData[Proc/out/H_\hn /mean/GARCH_\sn _norm_RES_mean_H_\hn .tex]{mytag}\\[8pt]}
\scriptsize{Note: The MSEs for $B,\Sigma, \Gamma$ are computed by stacking the elements of each matrix in a vector. $\mu_{\varepsilon}\equiv(\mu_{1},\hdots,\mu_{N})\T$, $\mu_{f}\equiv(\mu_{N+1},\mu_{N+2},\mu_{N+3})\T$, $\varphi_{\varepsilon}\equiv(\varphi_1,\hdots,\varphi_N)' $, $\varphi_{f}\equiv(\varphi_{N+1}, \varphi_{N+2}, \varphi_{N+3})'$, $\sigma_{\eta,\varepsilon}\equiv(\sigma_{\eta,1},\hdots,\sigma_{\eta,N})' $ and $\sigma_{\eta,f}\equiv({\sigma}_{\eta,N+1}, {\sigma}_{\eta,N+2},{\sigma}_{\eta,N+3})'$.}
		\end{table}
}
}
\FloatBarrier
\clearpage

\section{List of Ticker Symbols and Companies}\label{appendix7}
\centering
{\scriptsize
\begin{longtable}{llll}
\caption{List of the ticker symbols and names of the companies, ordered by market capitalization. The ordering is done column wise (the columns are continued on the next page) from top to bottom and from left to right. Thus JNJ has the largest market capitalization, XOM the second largest, and JWN the lowest.}\label{symbolnames}\\
\toprule
\endfirsthead
\toprule
\endhead
\bottomrule
\endfoot
\bottomrule
\endlastfoot
JNJ	    &    	Johnson \& Johnson		    			&    	VFC	    &    	VF Corp										\\
XOM	    &    	Exxon Mobil Corp		    			&    	XEL	    &    	Xcel Energy Inc								\\		
JPM	    &    	JPMorgan Chase \& Co		    		&    	STI	    &    	SunTrust Banks Inc							\\			
PG	    &    	Procter \& Gamble		    			&    	GLW	    &    	Corning Inc									\\	
BAC	    &    	Bank of Amarica Corp					&    	PPG	    &    	PPG Industries Inc							\\			
INTC	&    	Intel Corp		    					&    	ED	    &    	Consolidated Edison Inc					\\
CVX	    &    	Chevron Corp		    				&    	LUV	    &    	Southwest Airlines Co					\\					
PFE	    &    	Pfizer		    						&    	AMD	    &    	Advanced Micro Devices Inc				\\						
DIS	    &    	Walt Disney Co		    				&    	PH	    &    	Parker-Hannifin Corp					\\					
WFC	    &    	Wells Fargo \& Co		    			&    	MSI	    &    	Motorola Solutions Inc					\\
BA	    &    	Boeing Co		 	   	    			&    	PPL	    &    	PPL Corp								\\		
CMCSA	&    	Comcast Corp		    				&    	ES	    &    	Eversource Energy						\\				
KO	    &    	Coca-Cola Co		    				&    	ROK	    &    	Rockwell Automation Inc					\\					
PEP	    &    	PepsiCo Inc		    					&    	SWK	    &    	Stanley Black \& Decker Inc				\\						
MCD	    &    	McDonald's Corp		    				&    	EIX	    &    	Edison International					\\					
WMT	    &    	Walmart Inc		    					&    	WY	    &    	Weyerhaeuser Co							\\			
ABT	    &    	Abbott Laboratories		    			&    	HRS	    &    	Harris Corp								\\
IBM	    &    	International Business Machines Corp	&    	BLL	    &    	Ball Corporation						\\				
MMM	    &    	Minnesota Mining and Manufacturing Co	&    	KR	    &    	Kroger Co								\\		
UNP	    &    	Union Pacific Corp		    			&    	IP		&    	International Paper Co					\\					
MDT	    &    	Medtronic PLC		    				&    	NUE	    &    	Nucor Corp								\\		
HON	    &    	Honeywell International Inc		    	&    	ETR	    &    	Entergy Corp							\\			
TMO	    &    	Thermo Fisher Scientific Inc		    &    	OMC	    &    	Omnicom Group Inc						\\				
LLY	    &    	Eli Lilly \& Co		    				&    	HES	    &    	Hess Corp								\\		
UTX	    &    	United Technologies Corporation		    &    	HSY	    &    	Hershey Co								\\		
TXN	    &    	Texas Instruments Inc		    		&    	CMS	    &    	CMS Energy Corp							\\			
MO	    &    	Altria Group Inc		    			&    	CNP	    &    	CenterPoint Energy Inc					\\					
LOW	    &    	Lowe's Cos Inc		    				&    	WDC	    &    	Western Digital Corp					\\					
NEE	    &    	NextEra Energy Inc		   				&    	LEN	    &    	Lennar Corp								\\		
DHR	    &    	Danaher Corp		    				&    	K	    &    	Kellogg Co								\\		
GE	    &    	General Electric Co		    			&    	MYL	    &    	Mylan									\\	
CAT	    &    	Caterpillar Inc		    				&    	HST	    &    	Host Hotels \& Resorts Inc					\\					
AXP	    &    	American Express Co		 			    &    	MRO	    &    	Marathon Oil Corp							\\			
BMY	    &    	Bristol-Myers Squibb Co		    		&    	GWW	    &    	WW Grainger Inc								\\		
COP	    &    	ConocoPhillips		    				&    	DOV	    &    	Dover Corp									\\	
USB	    &    	US Bancorp		    					&    	CAG	    &    	Conarga Brands Inc							\\			
CVS	    &    	CVS Health Corp		    				&    	IFF	    &    	International Flavors \& Fragrances Inc	\\									
ADP	    &    	Automatic Data Processing Inc		    &    	L	    &    	Loews Corp									\\	
BDX	    &    	Becton Dickinson and Co		    		&    	CINF	&    	Cincinnati Financial Corp						\\				
TJX	    &    	TJX Cos Inc		    					&    	LNC	    &    	Lincoln National Corp							\\			
DUK	    &    	Duke Energy Corp		    			&    	APA	    &    	APA Corp										\\
D	    &    	Dominion Energy Inc		    			&    	NBL	    &    	Noble Energy Inc						\\				
SYK	    &    	Stryker Corp		    				&    	UDR	    &    	UDR Inc									\\	
CL	    &    	Colgate-Palmolive Co		    		&    	CMA	    &    	Comerica Inc								\\		
PNC	    &    	PNC Financial Services Group		    &    	TXT	    &    	Textron Inc									\\	
SO	    &    	Southern Co		    					&    	KSU	    &    	Kansas City Southern						\\				
SPGI	&    	S\&P Global Inc		    				&    	CTL	    &    	Lumen Technologies Inc						\\
DE	    &    	Deere \& Co							    &    	HRL	    &    	Hormel Foods Corp							\\			
RTN	    &    	Raytheon Co		    					&    	TAP	    &    	Molson Coors Beverage Co					\\
OXY	    &    	Occidental Petroleum Corp		    	&    	LNT	    &    	Alliant Energy Corp							\\			
EXC	    &    	Exelon Corp		    					&    	PKI	    &    	PerkinElmer Inc								\\		
MMC	    &    	Marsh \& McLennan Cos Inc		    	&    	NI	    &    	NiSource Inc								\\		
NOC	    &    	Northrop Grumman Corp		    		&    	FRT	    &    	Federal Realty Investment Trust				\\						
GD	    &    	General Dynamics Corp		    		&    	PNW	    &    	Pinnacle West Capital Corp					\\					
FDX	    &    	FedEx Corp		    					&    	HAS	    &    	Hasbro Inc									\\	
ECL	    &    	Ecolab Inc							    &    	PVH	    &    	PVH Corp									\\	
WBA	    &    	Walgreens Boots Alliance Inc		    &    	AVY	    &    	Avery Dennison Corp								\\		
TGT	    &    	Target Corp		    					&    	ZION	&    	Zions Bancorp NA								\\		
APD	    &    	Air Products and Chemicals Inc		    &    	SNA	    &    	Snap-on Inc										\\
PGR	    &    	Progressive Corp		   				&    	IPG	    &    	Interpublic Group of Cos Inc					\\					
KMB	    &    	Kimberly-Clark Corp		    			&    	SEE	    &    	Sealed Air Corp									\\	
ADI	    &    	Analog Devices Inc		    			&    	CPB	    &    	Campbell Soup Co								\\		
AEP	    &    	Americal Electric Power Co Inc		    &    	ALK	    &    	Alaska Air Group Inc							\\			
AMAT	&    	Applied Materials Inc		    		&    	FL	    &    	Foot Locker Inc									\\	
AIG	    &    	American International Group Inc	    &    	PNR	    &    	Pentair PLC										\\
BAX	    &    	Baxter International Inc		 	    &    	XRX	    &    	Xerox Holdings Corp								\\		
HUM	    &    	Humana Inc		    					&    	HP	    &    	Helmerich \& Payne Inc							\\			
AFL	    &    	Aflac Inc		    					&    	ROL	    &    	Rollins Inc										\\
F	    &    	Ford Motor Co		   			 		&    	NWL	    &    	Newell Brands Inc								\\		
ETN	    &    	Eaton Corp PLC		    				&    	GPS	    &    	Gap Inc											\\
WMB	    &    	Williams Cos Inc		    			&    	LEG	    &    	Leggett \& Platt Inc							\\			
SYY	    &    	Sysco Corp				 		 	 	&    	HRB	    &    	H\&R Block Inc									\\	
HPQ	    &    	HP Inc		    						&    	JEF	    &    	Jefferies Financial Group Inc					\\					
PEG	    &    	Public Service Enterprise Group Inc		&    	JWN	    &    	Nordstrom Inc								\\
\end{longtable}}		
\newpage

\section{Principal Component of the Real Data}\label{appendix7a}

\begin{figure}[ht!]\centering
\includegraphics[width=0.8\textwidth]{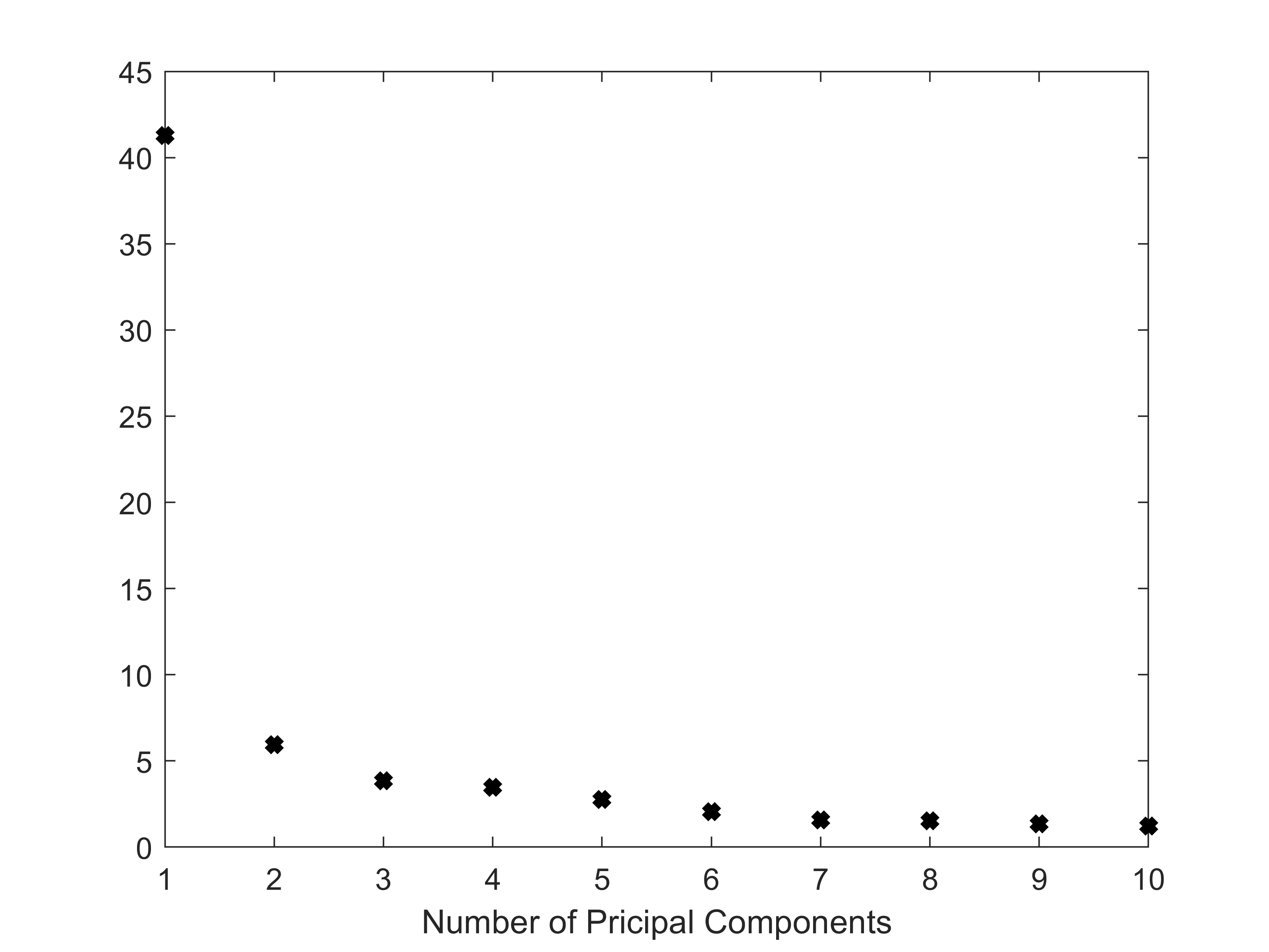}
\captionof{figure}{Plot of the largest ten sorted eigenvalues of the sample covariance matrix of the standardized returns of our empirical application.}\label{Eig_real}
\end{figure}
\FloatBarrier
\newpage

\section{Empirical Results for $k=1$}\label{appendix8}
\begin{figure}[ht!]
\centering
\includegraphics[width = .8\linewidth]{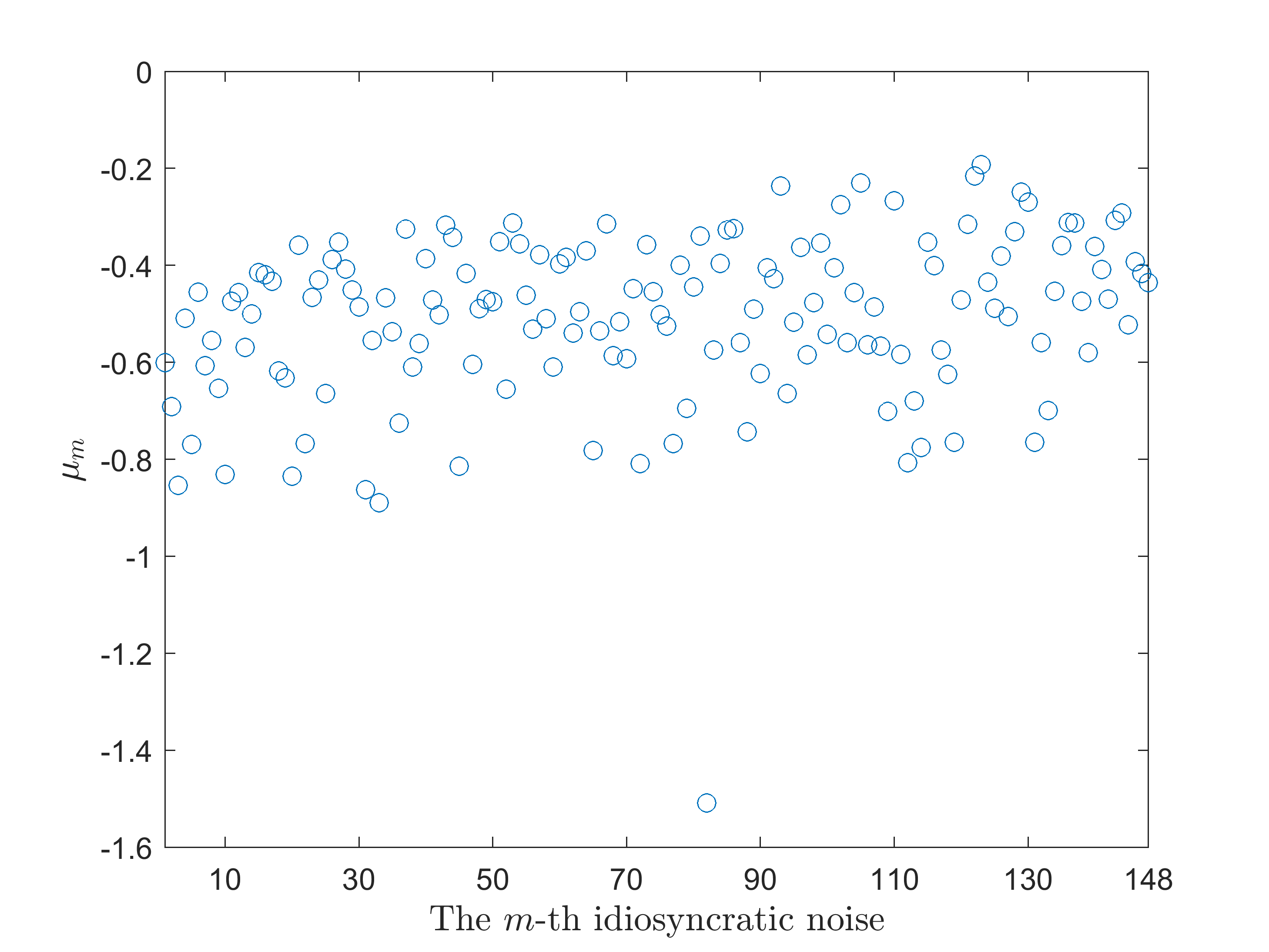}
\caption{Estimated constant of ARSV for each $m$-th idiosyncratic error, where $m=1,\ldots,148$ stocks}\label{real_mui_k1}
\end{figure}

\begin{figure}[ht!]
\centering
\includegraphics[width = .8\linewidth]{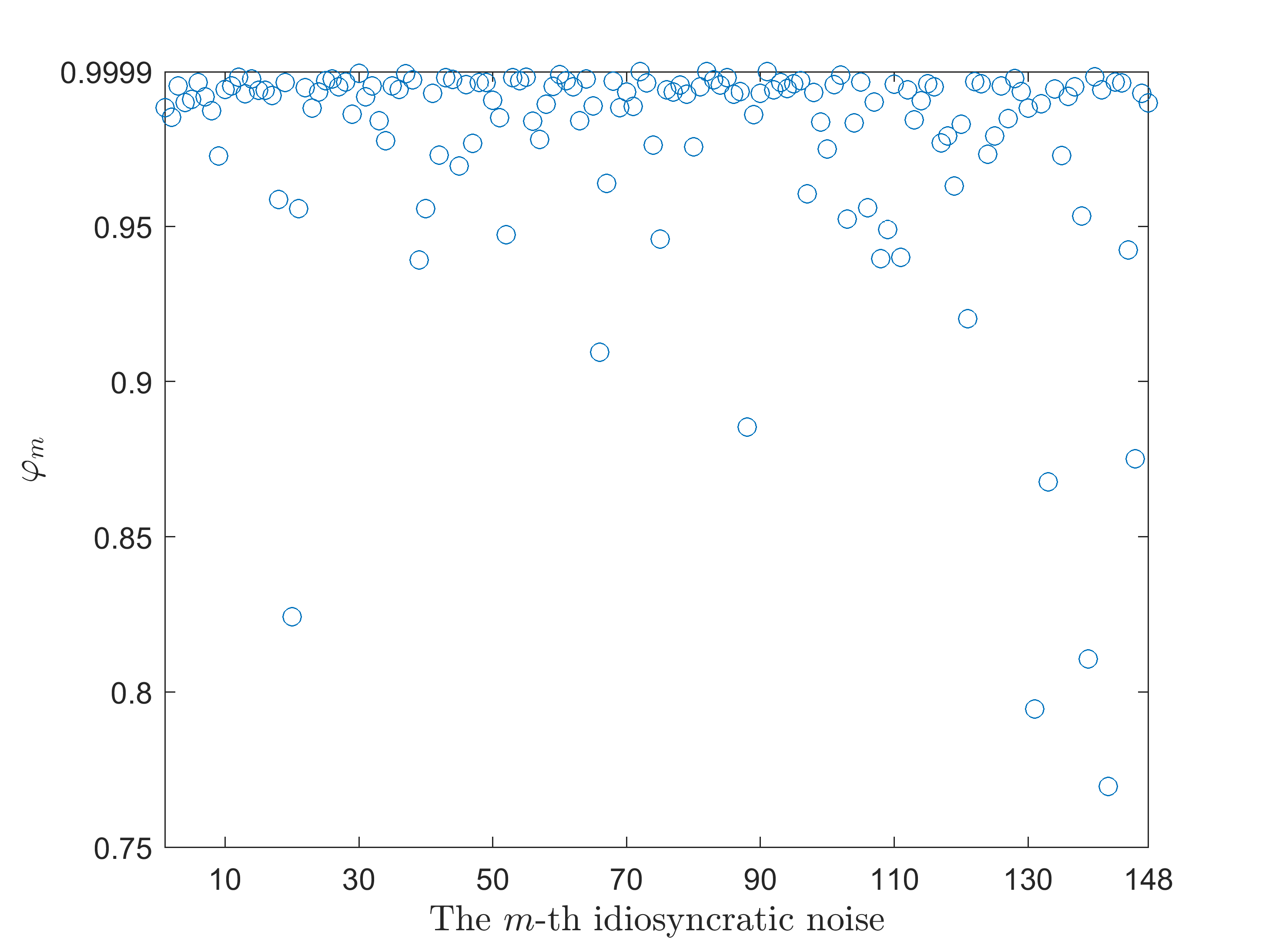}
\caption{Estimated autoregressive parameter of ARSV for each $m$-th idiosyncratic error, where $m=1,\ldots,148$ stocks}\label{real_phii_k1}
\end{figure}

\begin{figure}[ht!]
\centering
\includegraphics[width = .8\linewidth]{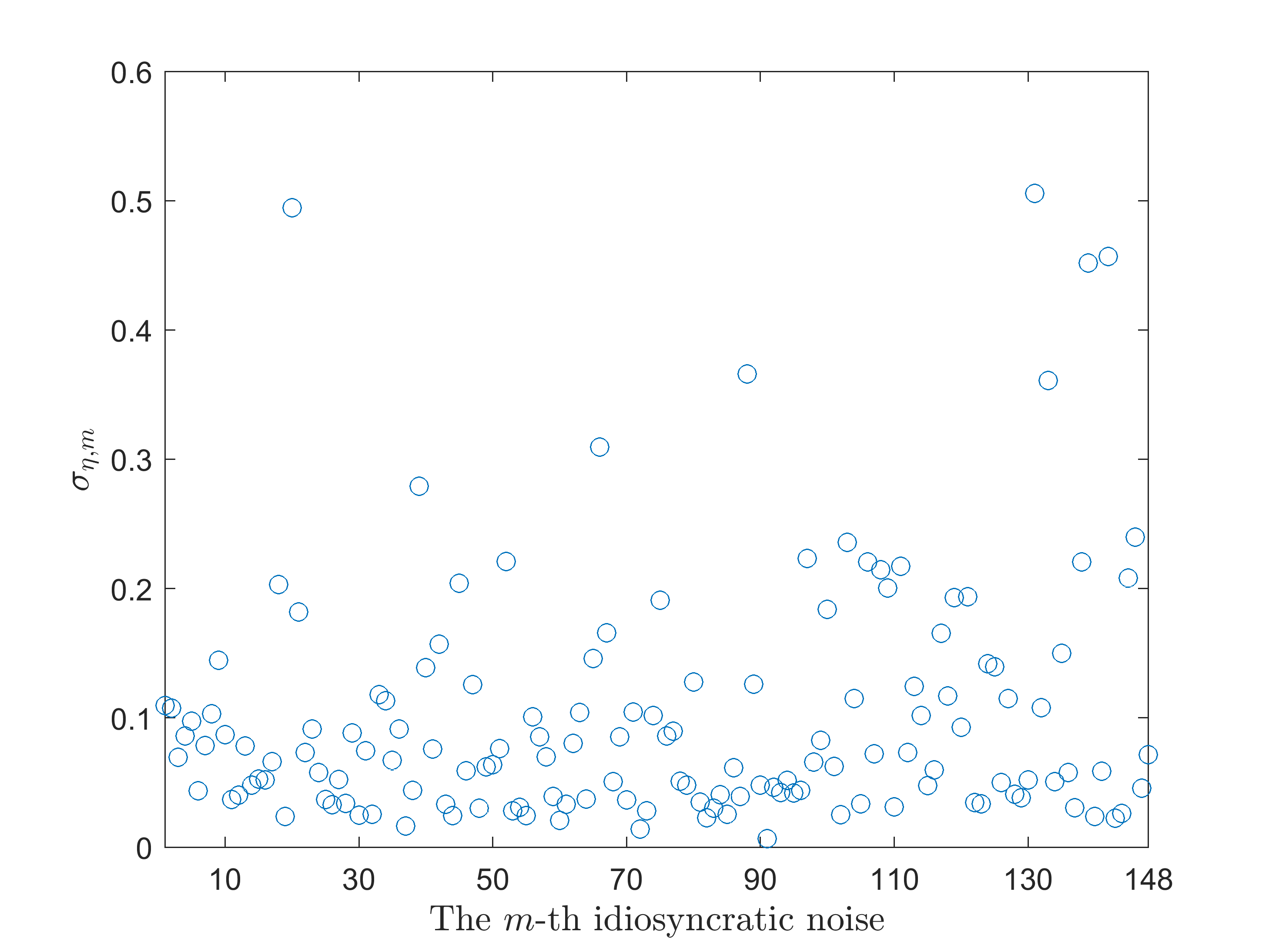}
\caption{Estimated standard deviation parameter of ARSV for each $m$-th idiosyncratic error, where $m=1,\ldots,148$ stocks}\label{real_setai_k1}
\end{figure}

\FloatBarrier
\clearpage
\section{Empirical Results for $k=1,2,3$}\label{appendix9}
\setlength{\columnsep}{0.16\linewidth}
\begin{table}[H]\centering
\caption{Estimates of the ARSV parameters of the $k$ factors. Standard errors in parentheses.}\label{real_factors}
\begin{tabular}{l r @{.} l r @{.} l r @{.} l | l r @{.} l r @{.} l r @{.} l }
\toprule
&\multicolumn{2}{c}{$k = 1$} & \multicolumn{2}{c}{$k = 2$} & \multicolumn{2}{c}{$k = 3$} & &
\multicolumn{2}{c}{$k = 1$} & \multicolumn{2}{c}{$k = 2$} & \multicolumn{2}{c}{$k = 3$}\\\midrule
${\gamma^{2}}_{N+1}$&    0&2918    &0&3042    &0&3466    &${\varphi}_{N+1}$    &0&9752    &0&9739    &0&9738\\
     &    (0&0303)    &(0&3533)    &(0&0513)    &        &(0&0043)    &(0&5210)    &(0&0083)\\
${\gamma^{2}}_{N+2}$&    \multicolumn{2}{c}{}&    0&0000    &0&0096    &${\varphi}_{N+2}$&    \multicolumn{2}{c}{}    &0&9876    &0&9913\\
     &    \multicolumn{2}{c}{}&    (0&0000)    &(0&0036)    &    &    \multicolumn{2}{c}{}    &(0&5847)    &(0&0029)\\
${\gamma^{2}}_{N+3}$&    \multicolumn{2}{c}{}&    \multicolumn{2}{c}{}&    0&1328    &${\varphi}_{N+3}$&    \multicolumn{2}{c}{}&    \multicolumn{2}{c}{}    &0&9854\\
     &    \multicolumn{2}{c}{}&    \multicolumn{2}{c}{}&    (0&0793)    &    &    \multicolumn{2}{c}{}&    \multicolumn{2}{c}{}    &(0&0068)\\
\midrule
${\mu}_{N+1}$&    -1&5866    &-1&5301    &-1&3825    &${\sigma}_{\eta,N+1}$    &0&1863    &0&1871    &0&1827\\
     &    (0&1156)    &(10&6368)    &(0&6025)    &        &(0&0164)    &(3&4561)    &(0&1696)\\
${\mu}_{N+2}$&    \multicolumn{2}{c}{}&    -15&5277    &-5&2052    &${\sigma}_{\eta,N+2}$&    \multicolumn{2}{c}{}    &0&1672    &0&1392\\
     &    \multicolumn{2}{c}{}&    (140&0762)    &(0&3620)    &    &    \multicolumn{2}{c}{}    &(9&4228)    &(0&0215)\\
${\mu}_{N+3}$&    \multicolumn{2}{c}{}&    \multicolumn{2}{c}{}&    -2&6855    &${\sigma}_{\eta,N+3}$&    \multicolumn{2}{c}{}&    \multicolumn{2}{c}{}    &0&1964\\
     &    \multicolumn{2}{c}{}&    \multicolumn{2}{c}{}&    (3&1305)    &    &    \multicolumn{2}{c}{}&    \multicolumn{2}{c}{}    &(0&4754)\\
\bottomrule
\end{tabular}

\end{table}

\begin{figure}[ht!]
\centering
\includegraphics[width = .8\linewidth]{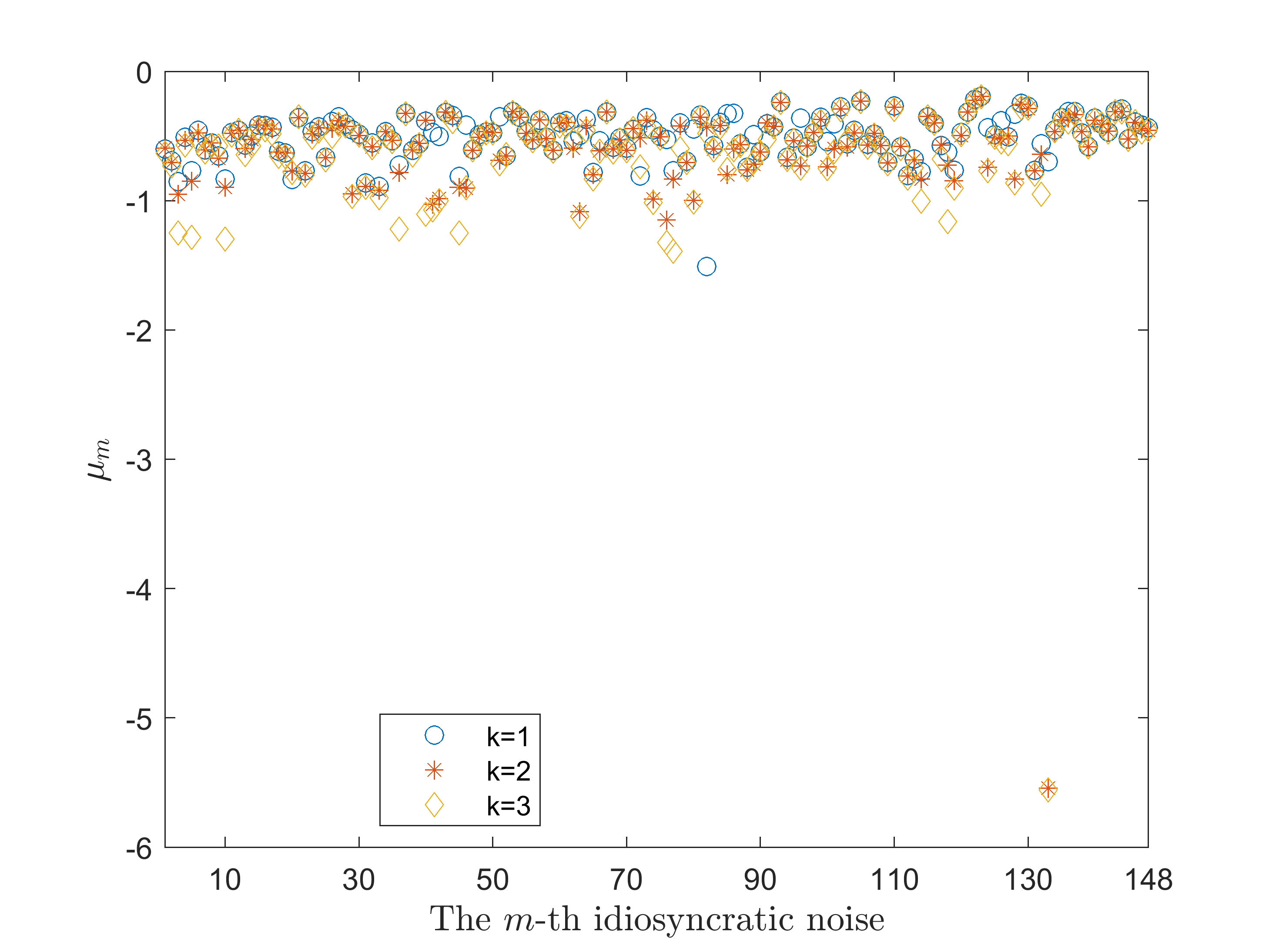}
\caption{Estimated constant of ARSV or the $m$-th idiosyncratic error, $m=1,\ldots,148$ stocks}\label{real_mui}
\end{figure}

\begin{figure}[ht!]
\centering
\includegraphics[width = .8\linewidth]{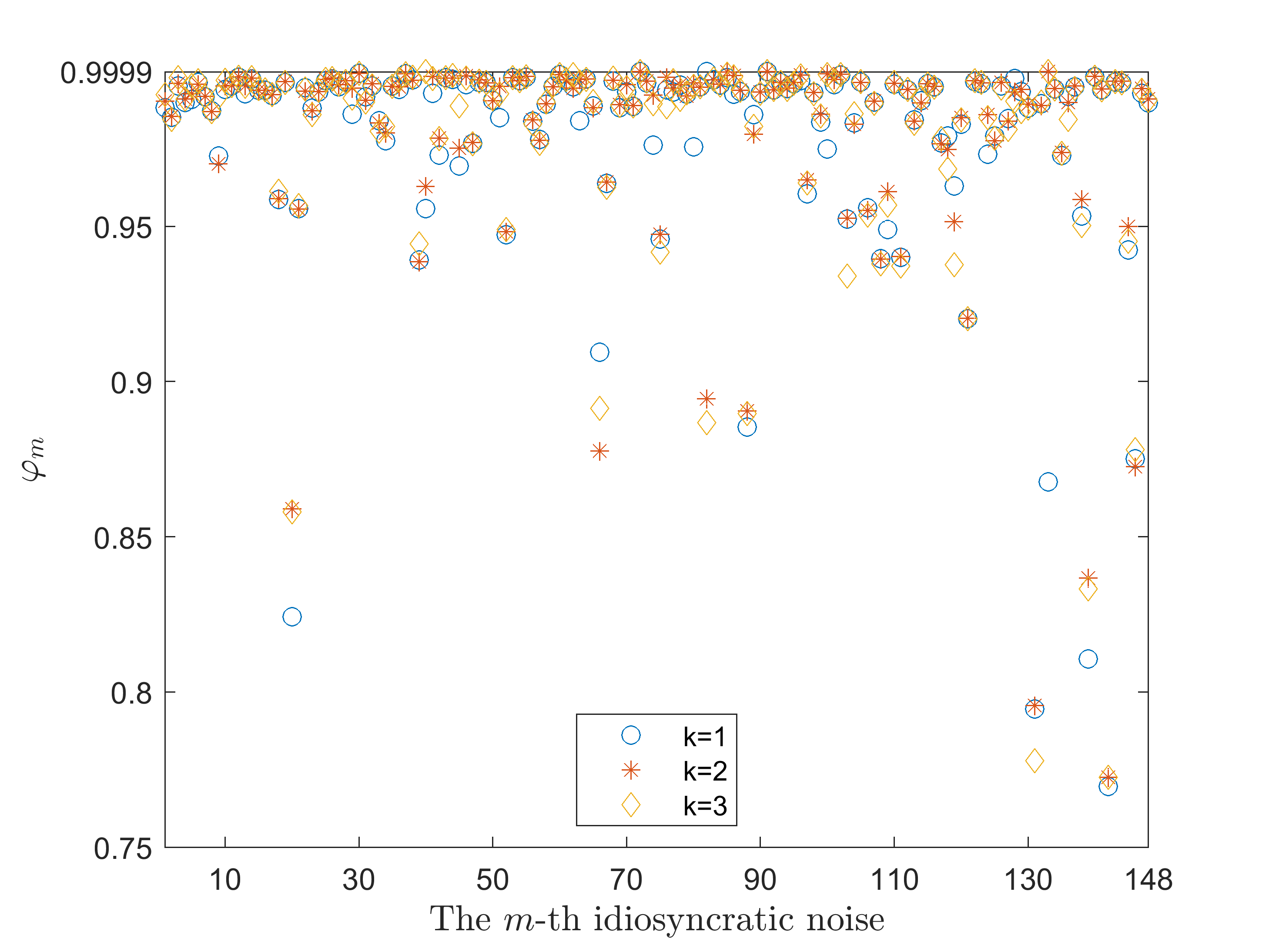}
\caption{Estimated autoregressive parameter of ARSV for the $m$-th idiosyncratic error, $m=1,\ldots,148$ stocks}\label{real_phii}
\end{figure}

\begin{figure}[ht!]
\centering
\includegraphics[width = .8\linewidth]{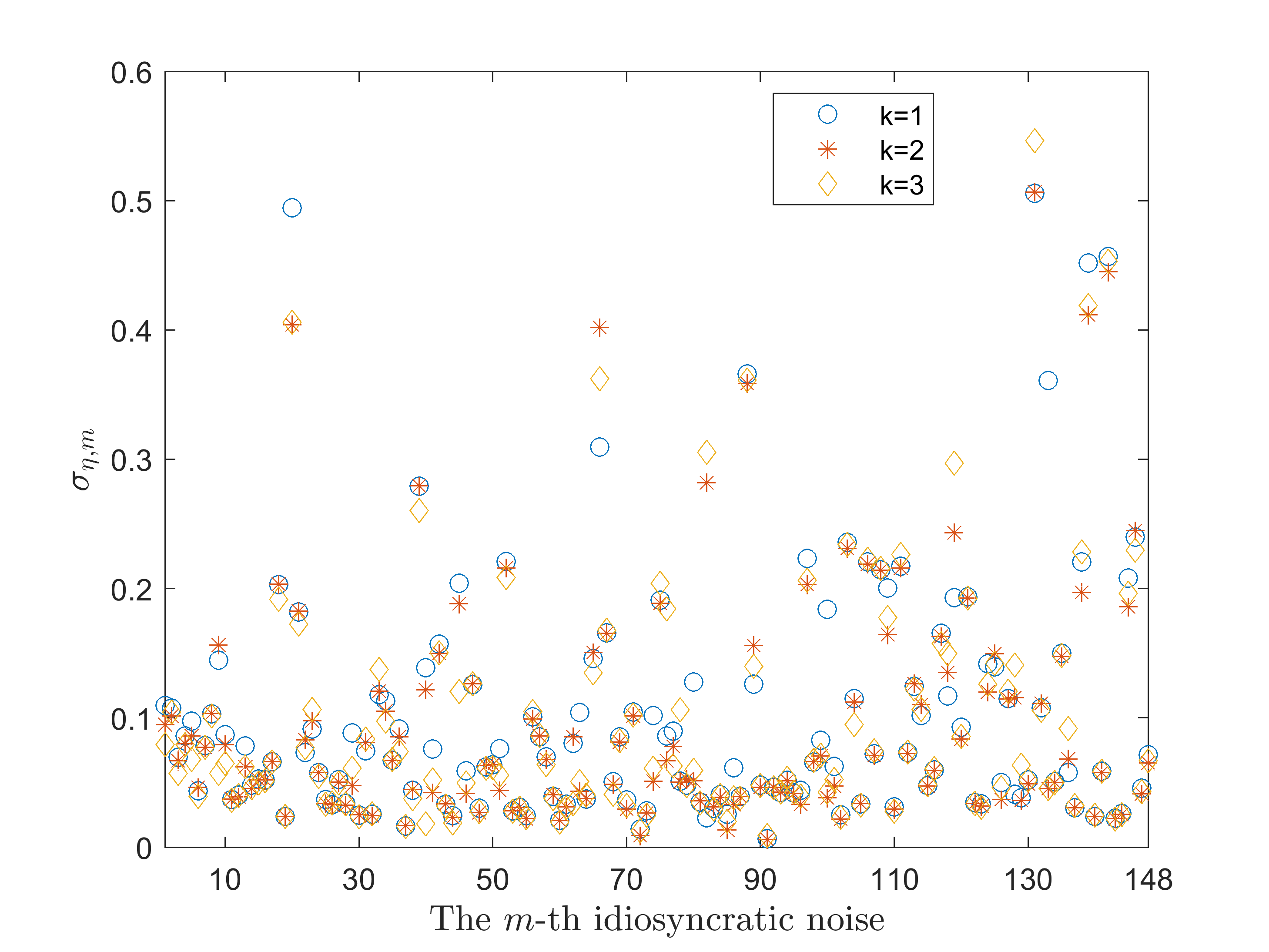}
\caption{Estimated standard deviation parameter of ARSV or the $m$-th idiosyncratic error, $m=1,\ldots,148$ stocks}\label{real_setai}
\end{figure}

\clearpage				
\begin{table}[h]\caption{Computational time of our approach and the Bayesian alternative in hours}\label{comptime}
\centering
\setlength{\columnsep}{0.1\linewidth}
\begin{tabular}{p{6cm} c r @{.} l r @{.} l r @{.} l}
\toprule
& \# CPUs	&\multicolumn{2}{c}{$\mathbf{k = 1}$} & \multicolumn{2}{c}{$\mathbf{k = 2}$} & \multicolumn{2}{c}{$\mathbf{k = 3}$} \\\midrule
\cite{kastner17}&1		&    58&9626  &  47&0225  &  51&2809\\\midrule
Our approach 	&1		&     3&5516  &  3&1101   &  4&3578\\
				&40 	&	  0&1558  &  0&1830   & 0&2327\\
\bottomrule
\end{tabular}
\end{table}

\begin{figure}[h]\centering
\includegraphics[width = 0.8\linewidth]{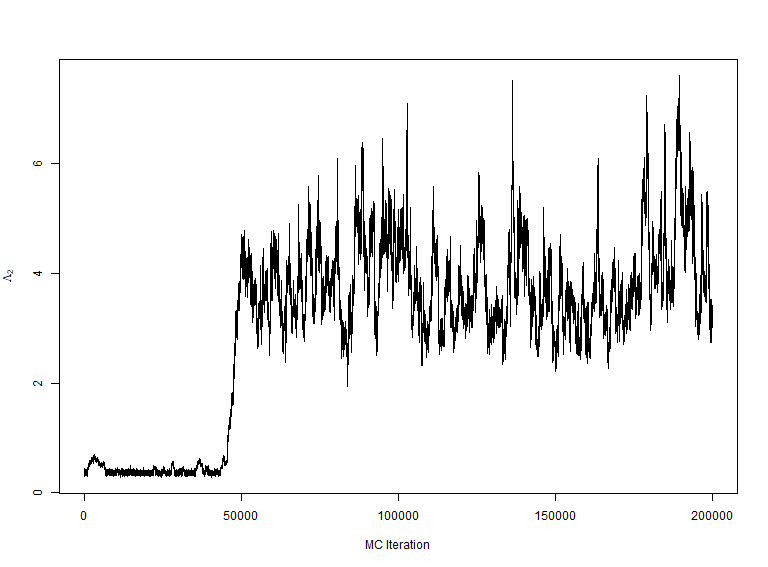}
\caption{Trace plot of the MCMC draws for the scaling parameter of the second factor for $k = 3$.}\label{kastner_trace}
\end{figure}

\begin{figure}[h]\centering
\includegraphics[width = 0.8\linewidth]{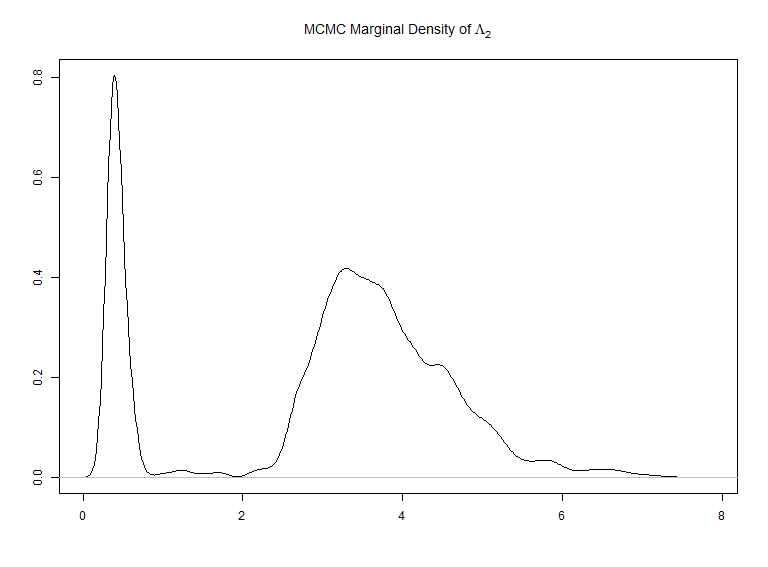}
\caption{Estimated MCMC density of the scaling parameter of the second factor for $k = 3$.}\label{kastner_dens}
\end{figure}

\FloatBarrier
\clearpage
	
\section{Monte Carlo Results on the Number of Factors}\label{appendix10}
\begin{figure}[ht!]
	\centering
\includegraphics[width=0.8\textwidth]{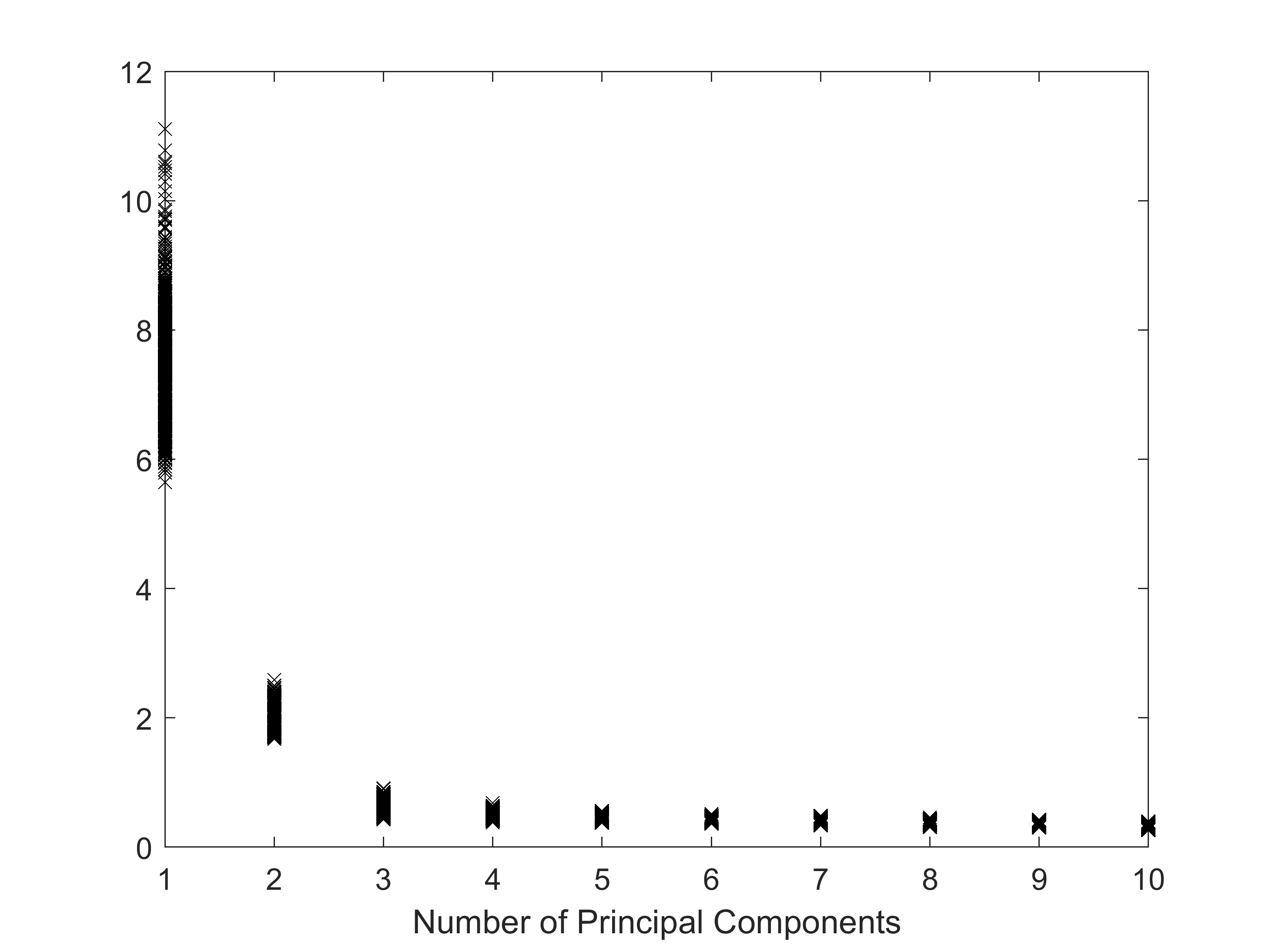}
\captionof{figure}{Plot of the sorted eigenvalues of the sample covariance matrix of the 1000 simulated  standardized return vectors for $N= 10$.}\label{Eig_N10}
\end{figure}

\begin{figure}[ht!]\centering
\includegraphics[width=0.8\textwidth]{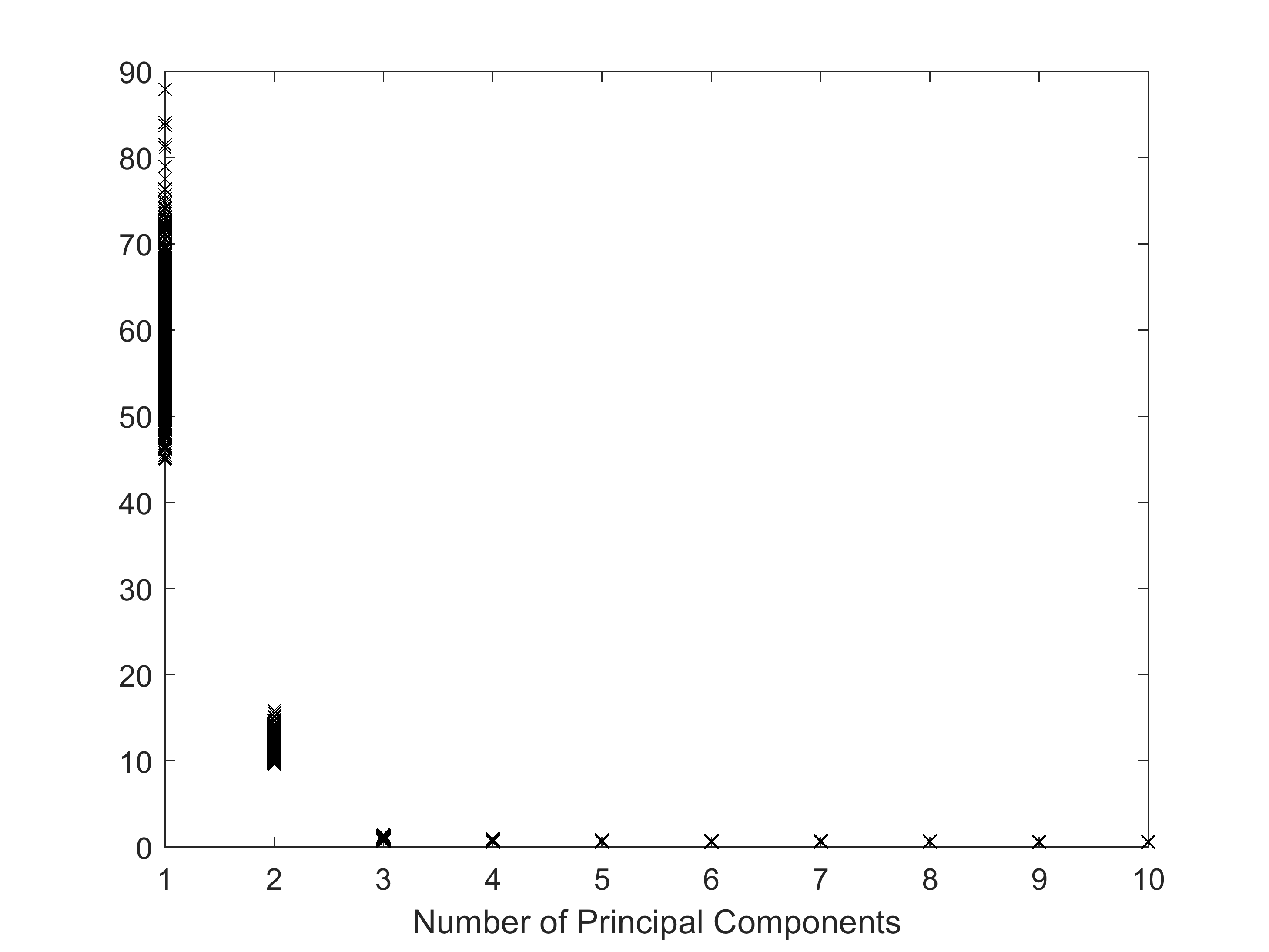}
\captionof{figure}{Plot of the largest ten sorted eigenvalues of the sample covariance matrix of the 1000 simulated standardized return vectors for $N= 100$.}\label{Eig_N100}
\end{figure}

\FloatBarrier
\clearpage
\begin{figure}[ht!]\centering
\includegraphics[width=0.8\textwidth]{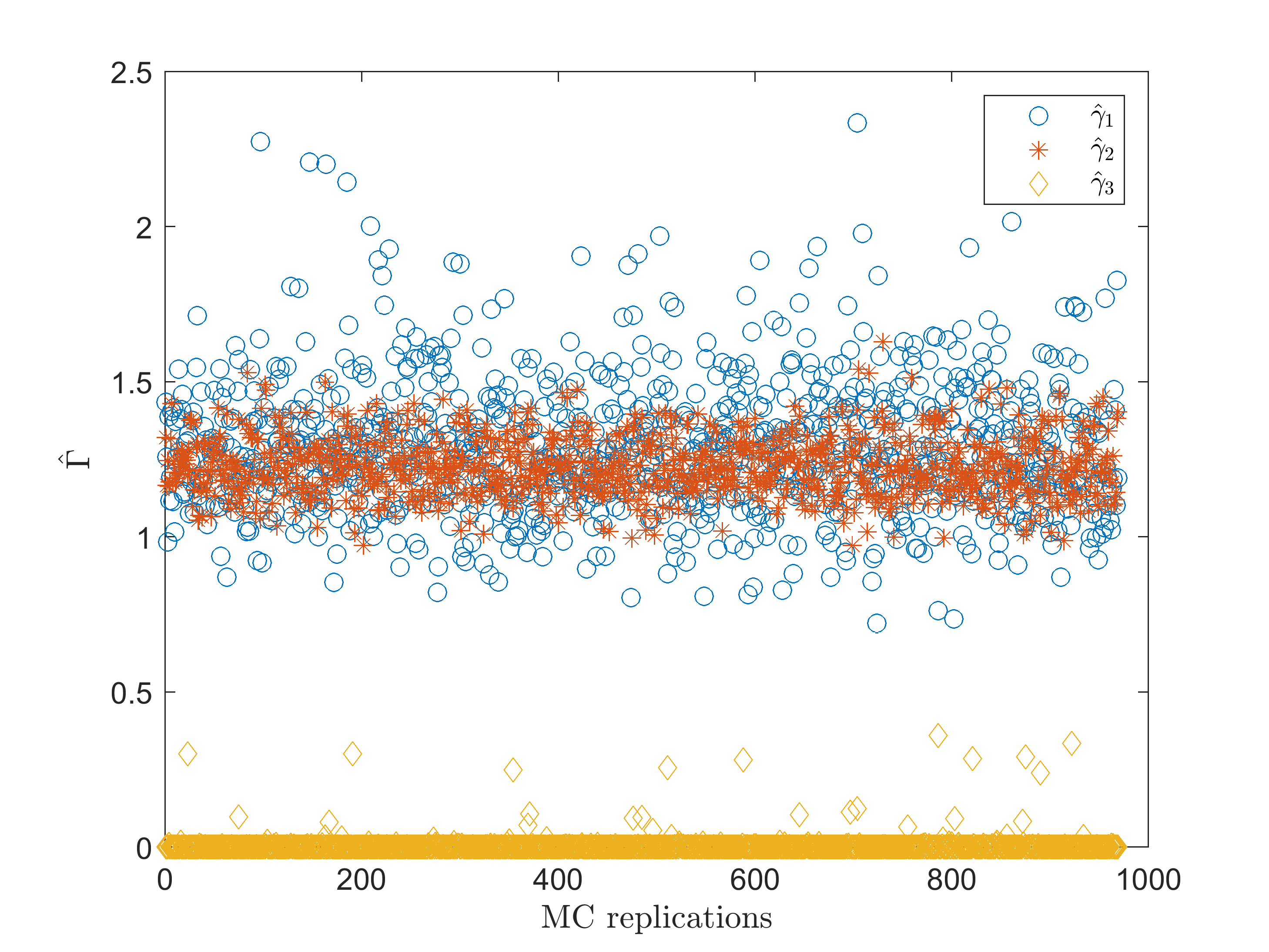}
\captionof{figure}{Plot of the estimated unconditional factor variances for $N= 10$ for the number of replications from 1 to 1000.}\label{gamma_N10}
\end{figure}

\begin{figure}[ht!]\centering
\includegraphics[width=0.8\textwidth]{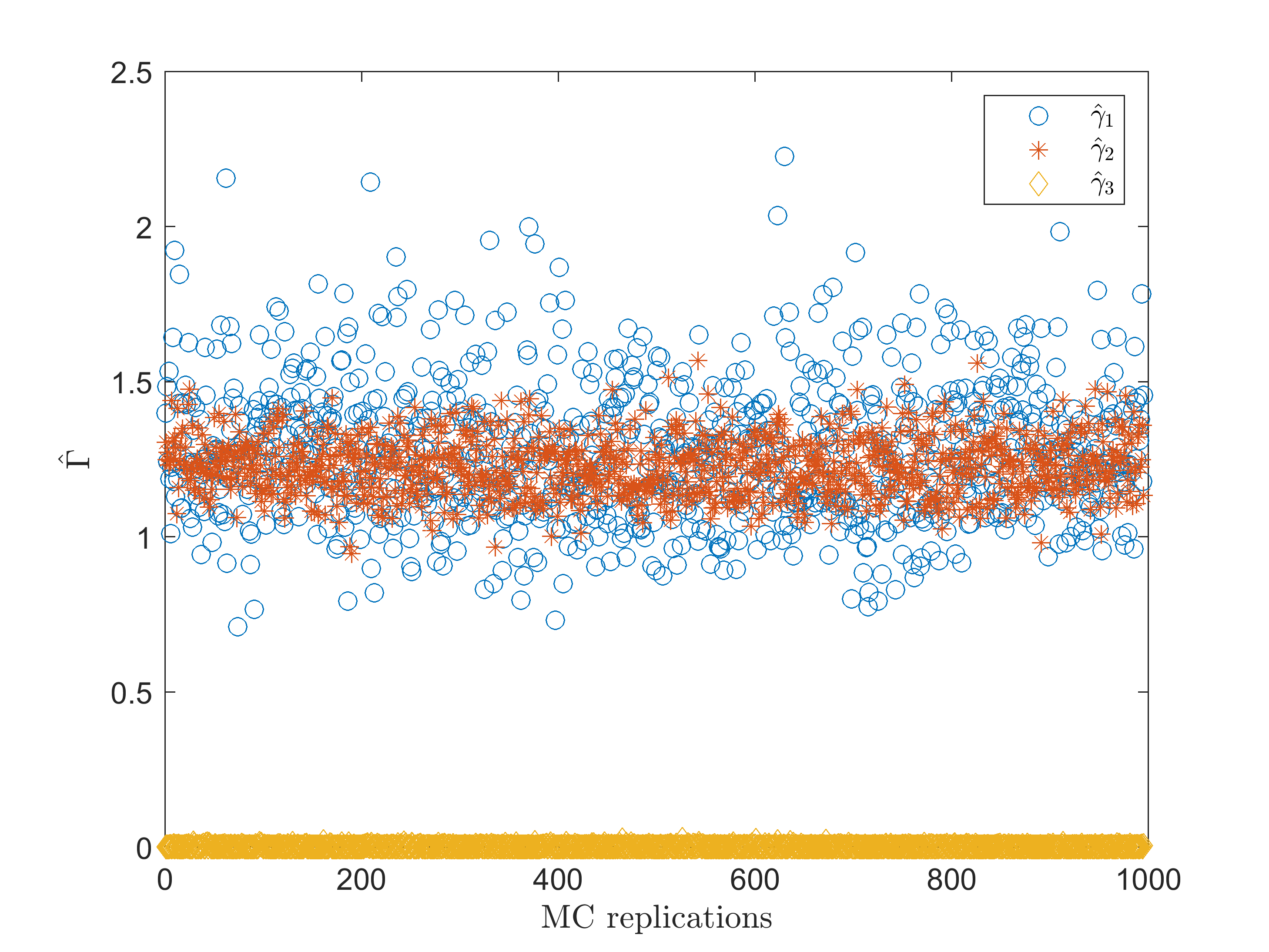}
\captionof{figure}{Plot of the estimated unconditional factor variances for $N= 100$
for the number of replications from 1 to 1000 }\label{gamma_N100}
\end{figure}

\end{document}